\documentclass[12pt]{article}

\usepackage{amsmath,amsthm,amssymb,a4wide}

\usepackage{mybbold}

\newcommand{\ud}{\text{d}}
\newcommand{\ui}{\text{i}}
\newcommand{\ue}{\text{e}}

\newcommand{\al}{\alpha}
\newcommand{\be}{\beta}
\newcommand{\de}{\delta}

\newcommand{\ga}{\gamma}

\newcommand{\la}{\lambda}

\newcommand{\om}{\omega}
\newcommand{\Om}{\Omega}

\newcommand{\vt}{\vartheta}

\newcommand{\ve}{\varepsilon}
\newcommand{\vp}{\varphi}

\newcommand{\cA}{{\mathcal A}}

\newcommand{\cD}{{\mathcal D}}

\newcommand{\cH}{{\mathcal H}}

\newcommand{\cS}{{\mathcal S}}

\newcommand{\vecA}{\vec{A}}

\newcommand{\vecB}{\vec{B}}

\newcommand{\vecC}{\vec{C}}
\newcommand{\vecE}{\vec{E}}
\newcommand{\vecx}{\vec{x}}
\newcommand{\vecX}{\vec{X}}
\newcommand{\vecy}{\vec{y}}
\newcommand{\vecz}{\vec{z}}
\newcommand{\vecp}{\vec{p}}
\newcommand{\vecP}{\vec{P}}
\newcommand{\vecR}{\vec{R}}
\newcommand{\vecs}{\vec{s}}
\newcommand{\vecsig}{\vec{\sigma}}
\newcommand{\vecSig}{\vec{\Sigma}}
\newcommand{\vecalph}{\vec{\alpha}}

\newcommand{\vecnab}{\vec{\nabla}}
\newcommand{\vecxi}{\vec{\xi}}

\newcommand{\rz}{{\mathbb R}}
\newcommand{\nz}{{\mathbb N}}
\newcommand{\gz}{{\mathbb Z}}
\newcommand{\kz}{{\mathbb C}}

\DeclareMathOperator{\tr}{Tr}
\DeclareMathOperator{\mtr}{tr}
\DeclareMathOperator{\im}{Im}
\DeclareMathOperator{\re}{Re}

\newcommand{\eins}{\mathmybb{1}}
\newcommand{\rto}{\rightarrow}

\numberwithin{equation}{section}

\theoremstyle{definition}

\theoremstyle{remark}

\newcommand{\kommentar}[1]{}

\begin{document}

\thispagestyle{empty}

\noindent
ULM-TP/98-6 \\
November 1998\\

\vspace*{1cm}

\begin{center}

{\LARGE\bf A semiclassical approach to the \\
\vspace*{3mm}
Dirac equation}%
{\Large
\footnote{This work has been submitted to Academic Press for possible 
          publication. Copyright may be transferred without notice, after 
          which this version may no longer be accessible.}} \\
\vspace*{2cm}
{\large Jens Bolte}%
\footnote{E-mail address: {\tt bol@physik.uni-ulm.de}} 
{\large and Stefan Keppeler}%
\footnote{E-mail address: {\tt kep@physik.uni-ulm.de}}\\ 

\vspace*{1cm}

Abteilung Theoretische Physik\\
Universit\"at Ulm, Albert-Einstein-Allee 11\\
D-89069 Ulm, Germany 
\end{center}

\vfill

\begin{abstract}
We derive a semiclassical time evolution kernel and a trace formula for the 
Dirac equation. The classical trajectories that enter the expressions are 
determined by the dynamics of relativistic point particles. We carefully 
investigate the transport of the spin degrees of freedom along the
trajectories which can be understood geometrically as parallel transport in a 
vector bundle with SU(2) holonomy. Furthermore, we give an interpretation in 
terms of a classical spin vector that is transported along the trajectories 
and whose dynamics, dictated by the equation of Thomas precession, gives rise 
to dynamical and geometric phases every orbit is weighted by. We also 
present an analogous approach to the Pauli equation which we analyse in two 
different limits.
\end{abstract}

\newpage

%%%%%%%%%%%% section 1%%%%%%%%%%%%%%

\section{Introduction}

\label{intro}

Most semiclassical approaches to the Dirac equation so far aimed at an
extension of the WKB method, with the expectation that some kind of 
Bohr-Sommerfeld quantisation conditions would emerge in relativistic 
quantum mechanics. The earliest such approach is due to Pauli 
\cite{Pau32}, who succeeded in showing that the phase of a WKB spinor 
is given by a solution of the Hamilton-Jacobi equation for relativistic 
point particles. But he could determine the amplitude of the semiclassical
spinor only in some special cases. Although the programme mentioned above has 
been very successful in nonrelativistic quantum mechanics, where it leads 
to the so-called Einstein-Brillouin-Keller (EBK) quantisation 
\cite{Kel58}, establishing semiclassical quantisation conditions in the 
case of the Dirac equation was found to be obstructed by the occurence 
of certain phases. Investigating multicomponent wave equations Yabana and 
Horiuchi \cite{YabHor87} noticed that geometric phases play an important 
r\^ole in this context and have to be incorporated in appropriate 
quantisation conditions. Using path integrals instead of the WKB method, 
Kuratsuji and Iida \cite{KurIid85,KurIid88} realised that an inclusion of the 
geometric phases in the symplectic form on classical phase space offers a 
possibility to arrive at quantisation conditions. A general theory of 
semiclassical quantisation for multicomponent wave equations in arbitrary
dimensions, which derive from classical Hamiltonian matrices with no 
globally degenerate eigenvalues, was developed by Littlejohn and Flynn 
\cite{LitFly91a,LitFly91b}. This method, however, does not apply to the 
Dirac Hamiltonian since the eigenvalues of the associated classical 
Hamiltonian matrix are twofold degenerate. Indeed, Emmrich and Weinstein
\cite{EmmWei96} found that in the degenerate case integrability of the 
corresponding classical dynamics is not a sufficiently strong condition 
to allow for an extension of EBK quantisation to the case of multicomponent 
wave equations. 

Moreover, even if successful, the procedure described above would not 
apply to systems whose classical limit is nonintegrable. In this paper we 
will therefore follow an alternative approach in that we investigate the 
semiclassical time evolution and then derive a semiclassical trace formula. 
This method is in the spirit of Gutzwiller's semiclassical treatment of 
the Schr\"odinger equation \cite{Gut71} (see also \cite{Gut90}), in which 
the quantum mechanical density of states is set into relation to a sum
over the periodic orbits of the corresponding classical system. The 
geometric phases mentioned above also appear in our approach in that they
represent the spin transport along classical orbits in the trace formula. 
The advantage of a trace formula approach is not only that it is applicable 
to both integrable and chaotic systems, but that it also provides the basis 
for efficient semiclassical quantisation conditions. In the context of 
quantum chaos extensive studies in this direction have been undertaken, 
see e.~g.\ \cite{Chaos2,AurMatSieSte92}. Furthermore, semiclassical 
trace formulae are the primary tools for a semiclassical theory of spectral 
statistics, see e.~g.\  \cite{BerTab77b,Ber85,BogKea96,Bol98}.

Before we go into more detail we want to take the opportunity to clarify
our point of view regarding the semiclassical limit on which we base the 
following investigations. In general we will consider the mathematical 
limit $\hbar\rto 0$, with the understanding that in a given physical 
situation an equivalent limit in terms of physical quantities such as 
controlable external parameters has to be taken. The effect of such a limit 
can then be expressed in terms of the spectrum of the Dirac Hamiltonian as 
follows. The l.h.s.\ of the trace formula that we are going to derive reads
\begin{equation}
\label{eq_lhs}
  \sum_n \varrho \left( \frac{E_n-E}{\hbar} \right) \ ,
\end{equation}
where the $E_n$'s are the quantum mechanical eigenvalues, which depend on
$\hbar$, and $E$ denotes a variable parameter. The smooth test function 
$\varrho$ decreases faster than any power for large arguments so that the 
main contribution to the sum comes from eigenvalues within an interval of 
length proportinal to $\hbar$,
\begin{equation}
\label{sclimit}
  E + \hbar \omega_1 < E_n < E + \hbar \omega_2 \ .
\end{equation}
In the formal limit $\hbar\rto 0$ the Weyl law forces the spectral density 
to increase in such a way that, although the length of the interval 
(\ref{sclimit}) shrinks to zero, infinitely many terms contribute to 
(\ref{eq_lhs}). In other words, the semiclassical limit corresponds to the 
limit of an increasing spectral density. The latter can possibly be achieved 
in a variety of different ways, which sometimes makes it necessary to vary 
several external parameters simultaneously. In case the limit $\hbar\rto 0$ 
is accompanied by further limits, one then has to ensure that asymptotic
expansions are uniform in the quantities that are involved in the 
further limits. Throughout this paper we will understand the 
semiclassical limit as being involved with the formal asymptotics 
as $\hbar\rto 0$. In the last section, where we investigate the 
nonrelativistic approximation of the results obtained for the Dirac 
equation, we briefly discuss an example with a second, simultaneous
limit. 

In the following sections we will develop the steps that are necessary
to derive a trace formula for the Dirac equation in some detail. Basically 
we follow the method introduced by Gutzwiller \cite{Gut67,Gut70,Gut71} in 
the case of the Schr\"odinger equation. It turns out that regarding the 
translational degrees of freedom Gutzwiller's approach can indeed be 
taken over. The novel features that arise in the case of the Dirac
equation derive from the spin degrees of freedom and their coupling 
to the translational dynamics. In section \ref{sec2} we first fix our
notation and recall the basic properties of the Dirac equation relevant 
for the following. Then we review the general r\^ole of periodic orbits 
in semiclassical trace formulae and introduce a regularisation procedure 
(cf.\ \cite{Mei92,PauUri95}) which allows to obtain convergent trace 
formulae. We moreover recall how to cut off the essential spectrum of the 
Dirac Hamiltonian, which is present in most physically relevant situations 
and typically covers $(-\infty,-mc^2] \cup [mc^2,\infty)$. 

Section \ref{sec3} is devoted to the derivation of a semiclassical time
evolution kernel in the spirit of the Van Vleck formula known for the
respective kernel for the Schr\"odinger equation. In the present context 
we find it convenient to represent the time evolution kernel in terms of 
an oscillatory integral. This method was developed for the study of scalar 
wave equations in the context of microlocal analysis, see e.~g.\ 
\cite{Dui96}, and subsequently found application to the development of 
several trace formulae \cite{CdV73a,DuiGui75,Mei92,PauUri95}. In the case 
of the Schr\"odinger equation it leads to the same result as Gutzwiller's 
original derivation \cite{Gut67} which employed a stationary phase
approximation of a Feynman path integral. The approach that we are going
to follow is similar to the usual WKB method and results in equations 
that determine the coefficients of an $\hbar$-expansion of the time 
evolution kernel. The presence of spin is reflected in these equations
through their matrix character. To lowest order one obtains as a 
condition for their solvability two Hamilton-Jacobi equations, which 
correspond to the classical dynamics of relativistic point particles with 
positive and negative kinetic energy, respectively. The condition that 
arises in next-to-leading order in $\hbar$ is usually called transport 
equation. The latter can be reduced to two differential equations for 
$2\times 2$ matrices describing the transport of the spin degrees of freedom 
along particle orbits. The solutions of these spin transport equations as 
well as the solutions of the Hamilton-Jacobi equations finally determine the 
leading order of the semiclassical expansion for the time evolution 
operator. It is remarkable that the classical Hamiltonians do not include 
any term corresponding to forces acting on the magnetic moment of the spin. 
This fact was already realised by Pauli in his WKB treatment of the Dirac 
equation, see also a related discussion in \cite{BalBlo74}. 
For this reason Pauli's method was subsequently 
criticized by de~Broglie \cite{Bro52}, who argued that electromagnetic 
moments linked with the spin were classical quantities and therefore 
should be present in a semiclassical approximation. This objection was 
later clarified by Rubinow and Keller \cite{RubKel63}. They pointed out 
that the moments of an electron are proportional to $\hbar$ so that in 
leading order as $\hbar \to 0$ the influence of spin on the trajectories 
vanishes. Furthermore, they showed how to obtain the equation describing 
the Thomas precession \cite{Tho27} of a classical spin, which is also known 
as the BMT-equation \cite{BarMicTel59}, from the transport equation. Since 
it only contains the ratio of the magnetic moment and the spin, $\hbar$ 
cancels from the equation of Thomas precession which therefore provides the 
correct description of a classical spin.

In section \ref{sec4} we analyse the spin transport in more detail, where 
we mainly focus on two aspects. We first discuss the geometric terms
that accompany semiclassical asymptotics of multicomponent wave equations
as they follow from the transport
equation. Their structure is of a similar form as discovered by Littlejohn 
and Flynn \cite{LitFly91b} in the case of wave equations with classical
Hamiltonian matrices that have no (globally) degenerate eigenvalues. In 
particular, one contribution is identified as being of the same type as 
the Berry phase \cite{Ber84,Sim83,ShaWil89} appearing in adiabatic 
approximations. However, since in the case of the Dirac equation the 
eigenvalues of the classical Hamiltonian matrix are twofold degenerate, 
the U(1)-holonomy factors of Littlejohn and Flynn are replaced by 
corresponding SU(2) terms. Our results are found to be in accordance with 
the general discussion of the transport equation for multicomponent wave 
equations by Emmrich and Weinstein \cite{EmmWei96}. In addition, these authors
revealed the global geometric meaning of all terms that contribute to the 
total holonomy following from the transport equation. As a second point we 
discuss how to express also the spin contributions to the semiclassical time 
evolution kernel in terms of classical quantities. To this end we introduce a 
classical spin vector $\vecs$ as an expectation value of a time dependent 
spin operator. It then follows from the spin transport equation that $\vecs$ 
has to fulfill the classical equation of Thomas precession. 
Thus, up to a phase factor the desired solution of the spin transport 
equation is determined by the classical spin precession along a given 
particle orbit. We also show that the additional phase factor is composed 
of a dynamical part associated with the energy of a classical magnetic 
moment in given electromagnetic fields, and a geometric part which is of 
the type discovered by Aharonov and Anandan \cite{AhaAna87}.

In section \ref{sec5} we derive the semiclassical trace formula for the 
Dirac equation, which is our central result, by Fourier transforming from 
the time domain to the energy domain and by subsequently taking the trace 
over spatial variables and matrix components. The periodic orbits that 
enter the trace formula are determined by the relativistic dynamics of 
classical point particles without internal degrees of freedom. The 
influence of the spin appears through two phase factors every periodic 
orbit is weighted by. One of these phases measures the change in the 
direction of the classical spin after this has been transported along a 
periodic orbit, whereas the second one contains the phase described above 
as being composed of a dynamical and a geometric part. 

Finally, in section \ref{sec6} we consider the nonrelativistic limit
of the results obtained in the previous sections. It turns out that the
leading order as $c\rto\infty$ coincides with the result of an application
of the above programme to the Pauli equation. This equation is also of 
independent interest because in applications it is often used to 
investigate spin-orbit coupling. Based on the method developed in 
\cite{LitFly91b}, this important effect is e.~g.\ treated semiclassically 
in \cite{LitFly92,FriGuh93}. Frisk and Guhr \cite{FriGuh93} introduced a 
trace formula for a nonrelativistic Hamiltonian that describes spin-orbit 
coupling by modifying the Berry-Tabor trace formula \cite{BerTab76}
for classically integrable systems without spin. To be able to use the 
formalism of \cite{LitFly91b} they kept Bohr's magneton 
$\mu=\frac{e\hbar}{2mc}$ fixed as $\hbar \to 0$; otherwise the corresponding 
classical Hamiltonian matrix would have a twofold degenerate eigenvalue.
This procedure is in contrast to our method which allows to treat $\hbar$ 
on the same footing in all terms. Fixing Bohr's magneton can be regarded as 
simultanously performing the limit $\hbar \to 0$ and the limit of an 
infinite coupling of spin to the translational degrees of freedom. In the 
semiclassical expressions the spin precession then decouples adiabatically 
from the translational motion. For the Pauli equation that describes a
coupling of spin to an external magnetic field we compare two ways of
performing the semiclassical limit: (i) $\hbar\to 0$ and (ii) $\hbar\to 0$,
$|\vecB|\to 0$ with $\hbar\,|\vecB|=const.$ In the first case the 
nonrelativisitic limit of the result for the Dirac equation emerges,
where the geometric terms in the spin transport are SU(2)-holonomy factors.
In the second case the adiabatic decoupling of the spin motion results
in geometric terms that yield a U(1) holonomy such that the Berry phase
of a precessing spin is recovered. Both results being different implies
that the semiclassical limit $\hbar \to 0$ is not uniform in the magnetic 
field strength, a result that sheds some light on de~Broglie's criticism
of Pauli's approach.

%%%%%%%%%%%%%%%%%% section 2 %%%%%%%%%%%%%%%%%

\section{Semiclassical asymptotics and classical trajectories}

\label{sec2}

In this section we will present the basis for our subsequent 
discussion of semiclassical methods for the Dirac equation. In the 
following our focus will be on relativistic particles of charge $e$
and mass $m$ with spin 1/2 in external static electromagnetic fields. 
Thus the relevant Dirac equation reads 
\begin{equation}
\label{Diraceqn}
\ui\hbar\,\frac{\partial\Psi}{\partial t}(\vecx,t)
= \hat H_D \Psi(\vecx,t) \ , 
\end{equation}
with the quantum Hamiltonian 
\begin{equation}
\label{DiracHam} 
\hat H_D := c\vecalph \left( \frac{\hbar}{\ui}\vecnab_{\vecx} - 
            \frac{e}{c} \vecA (\vecx) \right) 
            + \beta mc^2 + e \, \varphi(\vecx) \ ,
\end{equation}
which is a matrix-valued differential operator of first order. Here 
$\varphi$ and $\vecA$ denote the electromagnetic potentials such that the 
corresponding fields are given by $\vecE(\vecx)=-\vecnab_{\vecx}\varphi
(\vecx)$ and $\vecB(\vecx)=\vecnab_{\vecx}\times\vecA(\vecx)$. The Dirac 
algebra is realised by the $4\times 4$ matrices 
\begin{equation}
\vecalph = \left( \begin{array}{cc}  0 & \vecsig \\ 
                  \vecsig & 0 \end{array} \right)
\quad \text{and} \quad
\beta = \left( \begin{array}{cc}  \eins_{2}  & 0 \\
               0 & - \eins_{2}  \end{array} \right)\ ,
\end{equation}
where $\vecsig$ denotes the vector of Pauli matrices, and $\eins_{k}$ is 
the $k\times k$ unit matrix; see \cite{BjoDre64} for further details. 
The Hamiltonian (\ref{DiracHam}) can be realised as 
\begin{equation}
\hat H_D = H_D \left( \frac{\hbar}{\ui}\vecnab_{\vecx},\vecx \right)\ ,
\end{equation}
where 
\begin{equation}
\label{symbmat}
H_D \left( \vecp,\vecx \right) := 
c\vecalph \left( \vecp - \frac{e}{c} \vecA (\vecx) \right) 
+ \beta mc^2 + e \, \varphi(\vecx) 
\end{equation}
is the symbol matrix, in the sense of Weyl quantisation, of the operator 
$\hat H_D$ (see e.~g. \cite{Fol89}). In case the potentials $\varphi$ and 
$\vecA$ 
satisfy suitable regularity conditions, the Dirac Hamiltonian $\hat H_D$, 
when defined on the domain $C_0^\infty (\rz^3)\otimes\kz^4$ in the Hilbert 
space $\cH :=L^2 (\rz^3)\otimes\kz^4$, is essentially self-adjoint; see 
e.~g.\ \cite{Tha92,EvaLew97}. In the following we will always deal with 
its self-adjoint extension which we also denote by $\hat H_D$.

Solutions $\Psi (\vecx,t)$ of the Dirac equation (\ref{Diraceqn}), with
initial conditions $\Psi (\vecx,0)=\Psi_0 (\vecx)\in\cH$, are obtained 
as $\hat U(t)\Psi_0 =\ue^{-\frac{\ui}{\hbar}\hat H_D t}\Psi_0$. The time
evolution operator $\hat U(t)$ can then be represented by its Schwartz 
kernel $K(\vecx,\vecy,t)$, so that
\begin{equation}
\Psi(\vecx,t) = \int_{\rz^3} K(\vecx,\vecy,t) \, \Psi_0(\vecy) \ \ud^3 y \ .
\end{equation}
This matrix-valued kernel is obviously required to solve the Dirac 
equation, with initial condition
\begin{equation}
\label{incond}
\lim_{t\rto 0^+} K(\vecx,\vecy,t) = \eins_4\,\de (\vecx-\vecy) \ .
\end{equation}
Our first major goal, to be dealt with in the next section, will be to
derive a semiclassical representation of the time evolution kernel
$K(\vecx,\vecy,t)$ in the spirit of the Van~Vleck formula for the 
respective kernel of the Schr\"odinger equation. In the latter case such 
a representation is usually derived from a Feynman path integral, to
which the method of stationary phase is applied \cite{Gut67}. Here we
prefer an alternative semiclassical approach that makes use of a
representation of the kernel in terms of an oscillatory integral. This
method has previously also proven useful in the case of the Schr\"odinger
equation, see e.~g.\ \cite{Rob87,Bol98}, in which it served as a basis
for a mathematically rigorous proof \cite{Mei92,PauUri95} of the 
Gutzwiller trace formula \cite{Gut71,Gut90}.

We therefore now introduce the matrix-valued oscillatory integral
\begin{equation}
\label{oscillint}
K(\vecx,\vecy,t) = \frac{1}{(2\pi\hbar)^3} \int_{\rz^3}
                   \left[ a_\hbar^+ (\vecx,\vecy,t;\vecxi)\,
                   \ue^{\frac{\ui}{\hbar}\phi^+ (\vecx,\vecy,t;\vecxi)}
                   +a_\hbar^- (\vecx,\vecy,t;\vecxi)\,
                   \ue^{\frac{\ui}{\hbar}\phi^- (\vecx,\vecy,t;\vecxi)}
                   \right] \ \ud^3\xi\ ,
\end{equation}
where $\phi^\pm$ are real-valued smooth phase functions that are independent 
of $\hbar$, and $a_\hbar^\pm$ are $4\times 4$ matrices with semiclassical
expansions
\begin{equation}
\label{scamplitude}
a^\pm_\hbar (\vecx,\vecy,t;\vecxi) = \sum_{k=0}^{\infty} \left(
\frac{\hbar}{\ui}\right)^k \, a^\pm_k (\vecx,\vecy,t;\vecxi)
+O(\hbar^\infty) \ .
\end{equation}
According to the general theory of Schwartz kernels, $K(\vecx,\vecy,t)$
is a distribution kernel so that the integral (\ref{oscillint}) has to be 
interpreted in a distributional sense, see e.~g.\ \cite{Rob87} for details 
in the scalar case. In order to account for the initial condition 
(\ref{incond}) one chooses
\begin{equation}
\label{phiini}
\phi^\pm (\vecx,\vecy,0;\vecxi) =(\vecx - \vecy) \vecxi \ ,
\end{equation}
and
\begin{equation}
\label{ampini}
a_k^+ (\vecx,\vecy,0;\vecxi) + a_k^- (\vecx,\vecy,0;\vecxi) = 
\begin{cases} \eins_4 & \text{ if $k=0$ }\, , \\
              0 & \text{ if $k\geq 1$ }\, . \end{cases}
\end{equation}
Since the kernel (\ref{oscillint}) has to solve the Dirac equation, one
obtains conditions for the phases $\phi^\pm$ and the coefficients $a_k^\pm$
appearing in the matrix-valued amplitudes $a_\hbar^\pm$. A detailed
discussion of these equations will be postponed to the next section. Here
we only remark that the phase functions have to satisfy the Hamilton-Jacobi
equations
\begin{equation}
\label{HJG}
H^{\pm} \left( \vecnab_{\vecx} \phi^{\pm}(\vecx,\vecy,t;\vecxi),\vecx \right) 
+ \frac{\partial \phi^{\pm}}{\partial t}(\vecx,\vecy,t;\vecxi) = 0 \ ,
\end{equation}
with the classical Hamiltonians
\begin{equation}
\label{classHam}
H^{\pm}(\vecp,\vecx) = e\varphi(\vecx)\pm\sqrt{c^2\left(\vecp -\frac{e}{c}
                       \vecA(\vecx)\right)^2 + m^2c^4}
\end{equation}
of relativistic particles with positive $(+)$ and negative $(-)$ kinetic 
energies, respectively. As can readily be verified, the functions 
$H^{\pm}(\vecp,\vecx)$ are the two, twofold degenerate, eigenvalues of the 
symbol matrix (\ref{symbmat}). A posteriori, the occurrence of two 
Hamilton-Jacobi equations (\ref{HJG}) justifies the choice (\ref{oscillint}) 
of the oscillatory integral with two additive contributions. 

Due to the form (\ref{HJG}) of the Hamilton-Jacobi equations and the initial
conditions (\ref{phiini}), the variable $\vecy$ can be separated according
to 
\begin{equation}
\label{genfct}
\phi^{\pm}(\vecx,\vecy,t;\vecxi) = S^\pm (\vecx,\vecxi,t)-\vecy\vecxi\ ,
\end{equation}
so that the functions $S^\pm (\vecx,\vecxi,t)$ also solve the Hamilton-Jacobi 
equations (\ref{HJG}), but with initial conditions $S^\pm (\vecx,\vecxi,0)
=\vecx\vecxi$. From general Hamilton-Jacobi theory, see e.~g.\ 
\cite{Arn78,Rob87}, it is known that therefore $S^\pm (\vecx,\vecxi,t)$ 
are generating functions for canonical transformations 
$(\vecp,\vecx)\mapsto (\vecxi,\vecz)$. For ease of notation we restrict 
the following discussion to the index $+$. Then $(\vecp,\vecx)$ are the end 
points of the solution $(\vecP(t'),\vecX(t'))$ of Hamilton's equations of 
motion, generated by the Hamiltonian $H^+$, with initial condition
$(\vecxi,\vecz)$. This means that
\begin{equation}
(\vecP(0),\vecX(0)) = (\vecxi,\vecz) \quad \text{and} \quad 
(\vecP(t),\vecX(t)) = (\vecp,\vecx) \ .
\end{equation}
In order to explicitly incorporate the initial conditions we will also use
the notation
\begin{equation}
\label{classol}
\vecP(t') = \vecP(t';\vecxi,\vecz) \quad \text{and} \quad 
\vecX(t') = \vecX(t';\vecxi,\vecz) \quad \text{for} \quad 0\leq t'\leq t \ .
\end{equation}
The fact that $S^+ (\vecx,\vecxi,t)$ is a generating function for the
associated canonical transformation moreover implies the relations 
\begin{equation}
\label{generfct} 
\vecp = \vecnab_{\vecx} S^+ (\vecx,\vecxi,t) \quad \text{and} \quad
\vecz = \vecnab_{\vecxi} S^+ (\vecx,\vecxi,t) \ .
\end{equation}
For sufficiently small times, $|t|<t_c$, the Hamilton-Jacobi equations
(\ref{HJG}) are known to possess unique solutions $S^\pm(\vecx,\vecxi,t)$,
see e.~g. \cite{Rob87}. However, it is well known that at some critical
time $t_c$ a caustic may arise so that for $|t|\geq t_c$ the solutions of 
the Hamilton-Jacobi equations are no longer unique. The representation 
(\ref{oscillint}) of the time evolution kernel for $|t|\geq t_c$ can then 
only be a local one. A global object has to be constructed by gluing 
appropriate local representations together. This procedure requires 
consistency conditions, which are reflected by the presence of Morse indices 
in the final semiclassical expression to be derived in section \ref{sec3}. 
For the Schr\"odinger equation this was already noticed by Gutzwiller 
\cite{Gut67}. A mathematically rigorous construction employing the Maslov 
bundle can be found in \cite{Mei92}. For simplicity, we will neglect this 
problem below in that we continue to work with (\ref{oscillint}) and only 
in the end introduce the appropriate phase factors.  

Since the principal applications of semiclassical trace formulae deal with 
eigenvalues of quantum Hamiltonians, in particular with their semiclassical
determination and their statistical properties, respectively, one would
like to isolate the point spectrum of $\hat H_D$ from its essential
spectrum. In case the potentials $\vp$ and $\vecA$ vanish towards spatial 
infinity, or under suitably weakened conditions, the essential spectrum is 
known to be given by $\rz \backslash (-mc^2,mc^2)$, see e.~g.\ 
\cite{Tha92,EvaLew97}, so that for the following we assume $\hat H_D$ to 
have, possibly infinitely many, eigenvalues of finite multiplicities in 
the interval $(-mc^2,mc^2)$. Given any interval $I=(E_a,E_b)$ that contains 
only isolated eigenvalues, we introduce a smooth function $\chi(E)$ that 
is supported in $I$, i.~e., it vanishes outside of $I$, such that 
$\chi(E)=1$ on a suitably large subinterval. In particular, if there is no 
accumulation of eigenvalues at $E_a$ or $E_b$, one can achieve that 
$\chi (E_n)=1$ for all eigenvalues $E_n$ of $\hat H_D$. Otherwise, as e.~g.\ 
for the Dirac Hamiltonian of the hydrogen atom, one can enlarge the support 
of $\chi$ arbitrarily towards $\pm mc^2$ so that arbitrarily many 
eigenvalues can be taken into account. Then the operator $\chi(\hat H_D)$, 
defined by the functional calculus given by the spectral theorem, is 
bounded and self-adjoint; see e.~g.\ \cite{Rob87} for the scalar case. 
This operator has a purely discrete spectrum with eigenvalues $\chi (E_n)$. 
Therefore the Schwartz kernel $K_\chi (\vecx,\vecy,t)$ of the truncated 
time evolution operator $\hat U_\chi (t):=\chi(\hat H_D)\,\hat U(t)$ has 
a spectral representation
\begin{equation}
\label{Kspecrep}
K_\chi (\vecx,\vecy,t) = \sum_n \chi(E_n) \, \Psi_n(\vecx) 
\Psi_n^{\dag}(\vecy) \, \ue^{- \frac{\ui}{\hbar} E_n t}
\end{equation}
in terms of the orthonormal eigenspinors $\Psi_n$ of $\hat H_D$, where
$\Psi_n^{\dag}$ denotes the hermitian adjoint of $\Psi_n$. According to
the definition of the truncated time evolution operator, which implies that  
\begin{equation}
\left(\hat U_\chi(t)\Psi_0\right)(\vecx) = \chi(\hat H_D)\,\Psi(\vecx,t)\ ,
\end{equation}
the truncated kernel is obtained from the non-truncated one as
\begin{align}
K_\chi (\vecx,\vecy,t) 
   &= \chi(\hat H_D) \, K(\vecx,\vecy,t) \\
   &= \frac{1}{(2\pi\hbar)^3} \int_{\rz^3} \chi(\hat H_D) \left[ 
      a_\hbar^+ (\vecx,\vecy,t;\vecxi)\,
      \ue^{\frac{\ui}{\hbar}\phi^+ (\vecx,\vecy,t;\vecxi)}+
      a_\hbar^- (\vecx,\vecy,t;\vecxi)\,
      \ue^{\frac{\ui}{\hbar}\phi^- (\vecx,\vecy,t;\vecxi)}
      \right] \ \ud^3\xi\ ; \nonumber
\end{align}
here $\chi(\hat H_D)$ always acts on functions of $\vecx$. In the 
framework of Weyl calculus one can identify the symbol of the operator 
$\chi(\hat H_D)$ to possess a semiclassical expansion with principal 
symbol (leading term) $\chi(H_D (\vecp,\vecx))$, see \cite{Rob87}. Thus, 
an explicit calculation yields that
\begin{equation}
\ue^{-\frac{\ui}{\hbar}\phi(\vecx)} \left( \chi(\hat H_D) \, a \,
\ue^{\frac{\ui}{\hbar}\phi} \right) (\vecx) =
\chi \left(H_D(\vecnab_{\vecx} \phi(\vecx),\vecx)\right) \, a(\vecx)
+ O(\hbar) \ ,
\end{equation}
for any sufficiently regular matrix-valued function $a$ and real-valued 
function $\phi$. This calculation is closely parallel to the respective 
result in the scalar case, which can be found in \cite{Dui96}. Notice 
that in \cite{Dui96} a different quantisation is used which, however,
yields to lowest semiclassical order the same result as Weyl quantisation.
Thus, to leading order in $\hbar$, the truncated time evolution kernel 
reads
\begin{equation}
\label{Kchismcl} 
\begin{split}
K_\chi (\vecx,\vecy,t) = 
\frac{1}{(2\pi\hbar)^3} \int_{\rz^3}  
      & \left[ \chi (H_D(\vecnab_{\vecx}\phi^+ ,\vecx))\,a_0^+\,
        (1+O(\hbar)) \, \ue^{\frac{\ui}{\hbar}\phi^+}\right. \\
      & \ \left. + \chi (H_D(\vecnab_{\vecx}\phi^- ,\vecx))\,a_0^-\,
        (1+O(\hbar)) \, \ue^{\frac{\ui}{\hbar}\phi^-} 
        \right]\ \ud^3\xi \ .
\end{split}
\end{equation}

In order to prepare for the semiclassical trace formula to be dealt with
in section \ref{sec5} we now introduce a regularisation of the truncated
time evolution operator. To this end consider a smooth test function
$\varrho\in C^\infty (\rz)$ such that its Fourier transform
\begin{equation}
\tilde\varrho (t) := \int_{-\infty}^{+\infty} \varrho (E)\,\ue^{iEt}\ \ud E
\end{equation}
is smooth and compactly supported. Then in particular, both $\varrho$ and 
$\tilde\varrho$ are test functions from the Schwartz space $\cS(\rz)$. We now 
define the (bounded) operator
\begin{equation}
\hat U_\chi[\tilde\varrho] := \frac{1}{2\pi}\int_{-\infty}^{+\infty} 
\tilde\varrho(t)\,\hat U_\chi(t)\ \ud t \ ,
\end{equation}
whose trace can be calculated with the spectral representation 
(\ref{Kspecrep}) of the truncated time evolution  kernel as
\begin{equation}
\label{TrUchi}
\tr \hat U_\chi[\tilde\varrho] = 
\frac{1}{2\pi}\int_{\rz^3}\int_{-\infty}^{+\infty}\tilde\varrho(t)\,
\mtr K_\chi (\vecx,\vecx,t)\ \ud t\,\ud^3 x =
\sum_n \chi (E_n)\,\varrho\left(\frac{E_n}{\hbar}\right) \ .
\end{equation}
Here $\tr$ means the operator trace on $L^2 (\rz^3)\otimes\kz^4$, which
includes a trace over the matrix components; the latter is denoted by 
$\mtr$. The linear map $\tilde\varrho\mapsto\tr \hat U_\chi[\tilde\varrho]$ 
defines a tempered distribution, denoted by 
$\tr \hat U_\chi[\cdot]\in\cS'(\rz)$, if the sum on the r.h.s.\ of 
(\ref{TrUchi}) converges absolutely. To this end one requires that
\begin{equation}
\label{trconve}
\left| \chi (E_n)\,\varrho\left(\frac{E_n}{\hbar}\right) \right|
=O\left(\frac{1}{n^{1+\ve}}\right)\ ,\quad \ve >0\ ,\quad n\rto\infty\ .
\end{equation}
The simplest case of finitely many eigenvalues in the support of $\chi$
obviously poses no problem. If, however, eigenvalues accumulate at some
$E_{acc}\in [-mc^2, mc^2]$, the truncation $\chi$ has to be chosen such
that $\chi(E_{acc})=0$ and, moreover, $\chi(E)$ vanishes sufficiently
fast as $E\rto E_{acc}$ in order to fulfill (\ref{trconve}). We now 
evaluate the distribution $\tr \hat U_\chi[\cdot]$ on the test function 
$\tilde\varrho (t)\,\ue^{\frac{\ui}{\hbar}Et}$ and thus obtain the relation
\begin{equation}
\label{pretrace}
\sum_n \chi (E_n)\, \varrho \left( \frac{E_n -E}{\hbar} \right) = 
\frac{1}{2\pi} \int_{\rz^3} \int_{-\infty}^{+\infty} \tilde\varrho(t)\,
\ue^{\frac{\ui}{\hbar}Et} \,\mtr K_\chi (\vecx,\vecx,t)\ \ud t\,\ud^3 x \ .
\end{equation}
The semiclassical trace formula we are aiming at results, if for the 
truncated time evolution kernel in (\ref{pretrace}) one introduces a
semiclassical representation and calculates the integrals with the method
of stationary phase. The details of this procedure will be carried out in 
section \ref{sec5}. Here we only remark that upon introducing the 
representation (\ref{Kchismcl}) for the kernel, in leading semiclassical 
order one has to compute the integrals
\begin{equation}
\label{statphint}
\int_{\rz^3} \int_{\rz^3} \int_{-\infty}^{+\infty} 
\tilde\varrho(t) \, \chi (H_D(\vecnab_{\vecx}\phi^\pm (\vecx,\vecx,t;\vecxi),
\vecx)) \, a_0^\pm (\vecx,\vecx,t;\vecxi) \, \ue^{\frac{\ui}{\hbar}
[\phi^\pm (\vecx,\vecx,t;\vecxi)+Et]} \ \ud t\,\ud^3 x\,\ud^3 \xi \ .
\end{equation}
According to the method of stationary phase, see e.~g.\ \cite{Rob87,Dui96}, 
all contributions to (\ref{statphint}) that exceed $O(\hbar^\infty)$ as 
$\hbar\rto 0$ are determined by the stationary points, in the variables 
$(\vec\xi,\vecx,t)$, of the phase $\phi^\pm (\vecx,\vecx,t;\vecxi)+Et$. 
These stationary points are solutions of the equations
\begin{gather}
\left[ \vecnab_{\vecx} \phi^\pm (\vecx,\vecy,t;\vecxi) +
\vecnab_{\vecy} \phi^\pm (\vecx,\vecy,t;\vecxi) \right]_{\vecy=\vecx} = 0 \ , 
\nonumber  \\
\frac{\partial\phi^\pm}{\partial t}(\vecx,\vecx,t;\vecxi)+E = 0 \ , \\ 
\vecnab_{\vecxi} \phi^\pm (\vecx,\vecx,t;\vecxi) = 0 \ .  \nonumber
\end{gather}
If one now recalls the connection (\ref{genfct}) of the phase $\phi^\pm$ 
to the generating function $S^\pm$, one obtains the equivalent conditions
\begin{equation}
\label{statpoin}
\vecnab_{\vecxi} S^\pm (\vecx,\vecxi,t) = \vecx \quad , \quad
\vecnab_{\vecx} S^\pm (\vecx,\vecxi,t) = \vecxi \quad , \quad 
\frac{\partial S^\pm}{\partial t}(\vecx,\vecxi,t) = -E \ ,
\end{equation}
to be fulfilled by stationary points $(\vec\xi,\vecx,t)$. A comparison 
with the relations (\ref{generfct}) now yields the conditions $\vecxi
=\vecp$ and $\vecx=\vecz$, so that the stationary points determine
periodic solutions, with energy $E$, of Hamilton's equations of motion 
generated by the Hamiltonians (\ref{classHam}). We denote these periodic
orbits by $\ga_p^\pm$. Their periods, given by the $t$-components of 
the corresponding stationary points, are called $T_{\ga_p^\pm}$. Notice
that our requirement on the test function $\tilde\varrho$ to have compact
support implies that only periodic orbits up to some finite period, 
$|T_{\ga_p^\pm}|\leq T_{max}$, contribute. However, since the support of 
$\tilde\varrho$ can be made arbitrarily large one can manage to include as 
many periodic orbits as desired. As our first observation on the way towards 
a semiclassical trace formula we thus now conclude that, apart from terms 
of $O(\hbar^\infty)$, in the semiclassical limit $\hbar\rto 0$ all 
contributions to the l.h.s.\ of (\ref{pretrace}) are due to classical 
periodic orbits of energy $E$ of relativistic point particles. 

A further observation can be made with (\ref{statphint}) if the set of 
stationary points $(\vec\xi,\vecx,t)$ divides into a sequence $M_k$, 
$k\in\nz_0$, of $m_k$-dimensional smooth submanifolds of 
$\rz^3\times\rz^3\times\rz$ such that the phase $S^\pm (\vecx,\vecxi,t)
-\vecx\vecxi +Et$ is non-degenerate transversal to the manifolds $M_k$.
The latter condition means that the matrix of second derivatives of the phase
with respect to $(\vec\xi,\vecx,t)$ has constant rank $7-m_k$ when restricted
to $M_k$. The flow generated by $H^\pm$ is then called clean, see
\cite{Mei92,PauUri95} for further details. In this situation the set
of periods is known to be discrete in $\rz$ \cite{GuiUri89}. 
Thus the periods cannot accumulate at some finite value $T$. Under these 
conditions the method of stationary phase implies that each manifold 
$M_k$ yields a contribution
\begin{equation}
\label{genterm}
\hbar^{\frac{1-m_k}{2}}\,\cA_{M_k}(\hbar)\,\ue^{\frac{\ui}{\hbar}S_{M_k}} 
+O(\hbar^\infty)\ ,\quad \text{with} \quad \cA_{M_k}(\hbar)=\sum_{j=0}^\infty
\hbar^j\,\cA_{M_k,j}\ ,
\end{equation}
to (\ref{pretrace}). Here $S_{M_k}$ is the action of any periodic orbit
$\ga_p^\pm$ contained in $M_k$, which can be computed as the integral of 
the canonical one-form $\vecp\,\ud\vecx$ along any closed path on $M_k$.
The coefficients $\cA_{M_k,j}$ are independent of $\hbar$ and arise in the 
method of stationary phase as certain integrals over $M_k$. In the case
of an isolated periodic orbit $\ga_p^\pm$ the manifold $M_k$ is given by 
the points on the primitive periodic orbit associated with $\ga_p^\pm$,
and thus $m_k =1$. If the phase is non-degenerate transversal to an isolated 
periodic orbit, the latter is called non-degenerate. For this situation we 
will explicitly calculate $\cA_{M_k,0}$ in section \ref{sec5}. Another case 
that can be dealt with explicitly concerns the hypersurfaces
\begin{equation}
\Om_E^\pm := \left\{ (\vecp,\vecx);\ H^\pm (\vecp,\vecx)=E \right\}
\end{equation}
of constant energy $E$ in phase space. The points $(\vecp,\vecx)\in\Om_E^\pm$ 
are obviously periodic under the flows generated by $H^\pm$, with trivial 
period $T_0 =0$. Since according to the above non-degeneracy condition we
assume that $M_0^\pm =\Om_E^\pm\times\{0\}$ are (five dimensional) smooth 
manifolds, $E$ is required to be a regular value for the respective flows.
Moreover, the associated leading terms in (\ref{genterm}) are of the order 
$\hbar^{-2}$. The explicit calculation of $\cA_{M_0^\pm,0}$ will also be 
performed in section \ref{sec5}. As a final remark let us mention that in 
case the dynamics generated by $H^\pm$ are integrable, phase space foliates 
into three dimensional invariant tori such that the respective manifolds 
$M_k$ are given by the rational ones among these tori. Thus $m_k =3$ so that 
the leading terms in (\ref{genterm}) are of the order $\hbar^{-1}$.  

%%%%%%%%%%%%%% section 3 %%%%%%%%%%%%%%%

\section{Semiclassical time evolution}

\label{sec3}

In this section we will determine semiclassical representations for
the time evolution kernel $K(\vecx,\vecy,t)$ and its truncated 
version $K_\chi (\vecx,\vecy,t)$, respectively. In a first step we
derive necessary conditions that must be imposed on the matrix-valued
amplitudes $a_\hbar^\pm$ and on the real-valued phases $\phi^\pm$ 
in order that the oscillatory integral (\ref{oscillint}) be a possible
ansatz for the time evolution kernel. In a second step we then employ
the method of stationary phase to (\ref{oscillint}), and from the result
we obtain a semiclassical expression in the spirit of the Van~Vleck
formula for the respective kernel of the Schr\"odinger equation.

The requirement that (\ref{oscillint}), together with the expansion 
(\ref{scamplitude}), be a semiclassical representation of the kernel 
$K(\vecx,\vecy,t)$ to all orders in $\hbar$ means that the oscillatory 
integral shall fulfill the Dirac equation (\ref{Diraceqn}) up to terms 
$O(\hbar^\infty)$. We therefore act with $\hat H_D -\ui\hbar\frac{\partial}
{\partial t}$ on (\ref{oscillint}) after having introduced the expansion 
(\ref{scamplitude}), and then group terms of like orders in $\hbar$. The 
phases $\phi^\pm$ and coefficients $a_k^\pm$ now have to satisfy equations 
that result from demanding that all coefficients of powers $\hbar^k$, 
$k=0,1,2,\dots$, vanish.  To lowest order ($k=0$) one thus obtains the 
equation
\begin{equation}
\label{order0}
\begin{split}
&\ue^{\frac{\ui}{\hbar}\phi^+ (\vecx,\vecy,t;\vecxi)}\,
 \biggl[ H_D\Bigl(\vecnab_{\vecx}\phi^+ (\vecx,\vecy,t;\vecxi),\vecx \Bigr)
 +\frac{\partial\phi^+}{\partial t} (\vecx,\vecy,t;\vecxi) \biggr]
 a_0^+ (\vecx,\vecy,t;\vecxi) \\
&\hspace*{1cm}+\ue^{\frac{\ui}{\hbar}\phi^- (\vecx,\vecy,t;\vecxi)}\,
 \left[ H_D\left(\vecnab_{\vecx}\phi^- (\vecx,\vecy,t;\vecxi),\vecx \right)
 +\frac{\partial\phi^-}{\partial t} (\vecx,\vecy,t;\vecxi) \right]
 a_0^- (\vecx,\vecy,t;\vecxi) = 0 \ ,
\end{split}
\end{equation}
in which $H_D(\vecp,\vecx)$ denotes the symbol matrix (\ref{symbmat}) of 
the quantum Hamiltonian $\hat H_D$. For the following it turns out to be 
convenient to demand an individual vanishing of the terms with index $+$ 
and $-$, respectively. Since the two twofold degenerate eigenvalues of 
the hermitian $4\times 4$ symbol matrix $H_D(\vecp,\vecx)$ are given by 
$H^\pm (\vecp,\vecx)$, see (\ref{classHam}), the condition (\ref{order0}) 
can be fulfilled as soon as the matrices $a_0^\pm$ are suitably composed 
of eigenvectors of $H_D (\vecnab_{\vecx}\phi^\pm,\vecx)$. Upon 
diagonalising the symbol matrix $H_D(\vecp,\vecx)$ one obtains an 
orthonormal basis for $\kz^4$ (endowed with the canonical scalar product) 
that consists of eigenvectors $\{ e_1(\vecp,\vecx),e_2(\vecp,\vecx) \}$ 
with eigenvalue $H^+ (\vecp,\vecx)$, and eigenvectors 
$\{ f_1(\vecp,\vecx),f_2(\vecp,\vecx) \}$ with eigenvalue 
$H^- (\vecp,\vecx)$. We now define the $4\times 2$ matrices 
$V(\vecp,\vecx)$ and $W(\vecp,\vecx)$ whose two coloumns are given by 
the vectors $e_1(\vecp,\vecx),e_2(\vecp,\vecx)$ and 
$f_1(\vecp,\vecx),f_2(\vecp,\vecx)$, respectively. In explicit terms 
these matrices read
\begin{equation}\label{VWdef}
\begin{split}
V(\vecp,\vecx) &= \frac{1}{\sqrt{2\ve(\vecp,\vecx)[\ve(\vecp,\vecx)+mc^2]}}
                  \begin{pmatrix} [\ve(\vecp,\vecx)+mc^2]\,\eins_2 \\
                  {[c\vecp - e\vecA(\vecx)]\,\vecsig} \end{pmatrix}
                  \ ,  \\
W(\vecp,\vecx) &= \frac{1}{\sqrt{2\ve(\vecp,\vecx)[\ve(\vecp,\vecx)+mc^2]}}
                  \begin{pmatrix} {[c\vecp - e\vecA(\vecx)]\,\vecsig} \\
                  -[\ve(\vecp,\vecx)+mc^2]\,\eins_2 \end{pmatrix} \ ,
\end{split}
\end{equation}
where
\begin{equation}
\ve(\vecp,\vecx) := \sqrt{\left(c\,\vecp-e\,\vecA(\vecx)\right)^2 
+m^2c^4} = H^+ (\vecp,\vecx)-e\vp(\vecx)\ .
\end{equation}
Since the eigenvectors $\{ e_1,e_2,f_1,f_2 \}$ are chosen to be orthonormal 
and form a basis for $\kz^4$, one obtains the relations
\begin{equation}
V^\dagger V = \eins_2\ ,\quad W^\dagger W =\eins_2\ ,\quad 
V\,V^\dagger + W\,W^\dagger = \eins_4\ .
\end{equation}
Moreover, the projectors $P_\pm (\vecp,\vecx)$ onto the eigenspaces
corresponding to the eigenvalues $H^\pm (\vecp,\vecx)$, respectively, are
given by $P_+ =V\,V^\dagger$ and $P_- =W\,W^\dagger$.

Following (\ref{genfct}), we now choose the generating functions $S^\pm$
to replace the phase functions $\phi^\pm$. If one then introduces suitable 
$2\times 4$ matrices $\widetilde V$ and $\widetilde W$,
\begin{align}
\label{a0pm}
a_0^+ (\vecx,\vecy,t;\vecxi) 
&= V\left( \vecnab_{\vecx}S^+ (\vecx,\vecxi,t),\vecx \right)\,
   \widetilde V(\vecx,\vecxi,t) \ , \nonumber \\
a_0^- (\vecx,\vecy,t;\vecxi) 
&= W\left( \vecnab_{\vecx}S^- (\vecx,\vecxi,t),\vecx \right)\,
   \widetilde W(\vecx,\vecxi,t) \ ,
\end{align}
the condition (\ref{order0}) is fulfilled as soon as $S^\pm$, and hence 
also the phases $\phi^\pm$, satisfy the Hamilton-Jacobi equations
\begin{equation}
\label{HJGS}
H^{\pm} \left( \vecnab_{\vecx}S^{\pm}(\vecx,\vecxi,t),\vecx \right) 
+ \frac{\partial S^{\pm}}{\partial t}(\vecx,\vecxi,t) = 0 \ .
\end{equation}
In the definition of the matrices $\widetilde V$ and $\widetilde W$ we
anticipated the fact that the coefficients $a_k^\pm$ are independent of 
$\vecy$, see the discussion below (\ref{orderk}). At the moment an explicit 
expression for $\widetilde V$ and $\widetilde W$ is not needed. After having 
applied the method of stationary phase to (\ref{oscillint}) we will specify 
them further. Here we only remark that in order to fulfill the initial 
condition (\ref{ampini}) we demand that
\begin{equation}
\widetilde V (\vecx,\vecxi,0) = V^\dagger (\vecxi,\vecx) \quad
\text{and} \quad \widetilde W(\vecx,\vecxi,0) = 
W^\dagger(\vecxi,\vecx) \ ,
\end{equation}
so that $a_0^\pm (\vecx,\vecy,0;\vecxi)=P_\pm (\vecxi,\vecx)$. In 
conclusion, the condition (\ref{order0}) to lowest order in $\hbar$
requires the phases $\phi^\pm$ to solve the Hamilton-Jacobi equations
(\ref{HJG}). Due to the initial condition (\ref{phiini}) these functions
are therefore now completely fixed. 

Our next goal is to determine the leading contributions $a_0^\pm$ to the 
amplitudes. The condition (\ref{order0}) appearing to lowest order in
$\hbar$ only requires these coefficients to be of a certain structure, see 
(\ref{a0pm}). They are, however, completely fixed by the conditions imposed
in next-to-leading order. Regarding all higher orders $\hbar^k$, 
$k=1,2,3,\dots$, one observes that the equations obtained from the procedure 
described before (\ref{order0}) can be given in a uniform manner. They read
\begin{equation}
\label{orderk}
\begin{split}
 \biggl[ H_D \bigl(\vecnab_{\vecx}S^+,\vecx\bigr)+\frac{\partial S^+}
 {\partial t}\biggr] a_k^+ (\vecx,\vecy,t;\vecxi) &+ \biggl[ 
 c\,\vecalph\vecnab_{\vecx}+\frac{\partial}{\partial t} \biggr] 
 a_{k-1}^+ (\vecx,\vecy,t;\vecxi) = 0 \ ,\\
 \biggl[ H_D \bigl(\vecnab_{\vecx}S^-,\vecx\bigr)+\frac{\partial S^-}
 {\partial t}\biggr] a_k^- (\vecx,\vecy,t;\vecxi) &+ \biggl[ c\,
 \vecalph\vecnab_{\vecx}+\frac{\partial}{\partial t} \biggr] 
 a_{k-1}^- (\vecx,\vecy,t;\vecxi) = 0\ .
\end{split}
\end{equation}
Since these equations, as well as the initial conditions (\ref{ampini}),
are independent of $\vecy$, the coefficients $a_k^\pm$ do not depend
on $\vecy$ either. Beginning with $k=1$, the hierarchy (\ref{orderk}) of 
equations can principally be solved order by order. To this end, for each 
$k$ we multiply the two equations with $V^\dagger (\vecnab_{\vecx}S^+,\vecx)$ 
and $W^\dagger (\vecnab_{\vecx}S^-,\vecx)$, respectively, from the left. 
Since $V^\dagger H_D =V^\dagger H^+$ and  $W^\dagger H_D =W^\dagger H^-$, 
the Hamilton-Jacobi equations (\ref{HJGS}) then imply that
\begin{equation}        
\begin{split}
\label{orderkred}
V^\dagger (\vecnab_{\vecx}S^+,\vecx) \left[ c\,\vecalph\vecnab_{\vecx}
 +\frac{\partial}{\partial t} \right] a_{k-1}^+ (\vecx,\vecy,t;\vecxi)
& =0 \ , \\
W^\dagger (\vecnab_{\vecx}S^-,\vecx) \left[ c\,\vecalph\vecnab_{\vecx}
 +\frac{\partial}{\partial t} \right] a_{k-1}^- (\vecx,\vecy,t;\vecxi)
& =0 \ . 
\end{split}
\end{equation}
If one started with $a_0^\pm$ as given in (\ref{a0pm}), one now could 
in principle determine all higher coefficients $a_k^\pm$ from 
(\ref{orderk}) and (\ref{orderkred}) recursively. Our ambition is, however, 
limited to obtain a semiclassical expression for the time evolution kernel 
to leading order. In the following we therefore restrict our attention to 
the case $k=1$ and, moreover, only present the case with index $+$ 
explicitly.

Expressing $a_0^+$ as indicated in (\ref{a0pm}), the l.h.s.\ of the 
equation (\ref{orderkred}) for $k=1$ and index $+$ can be viewed as 
an application of the matrix-valued differential operator
\begin{equation}
\label{trans1}
V^\dagger (\vecnab_{\vecx}S^+,\vecx) \left[ c\,\vecalph\vecnab_{\vecx}
+\frac{\partial}{\partial t} \right] V(\vecnab_{\vecx}S^+,\vecx)
\end{equation}
to the matrix-valued function $\widetilde V(\vecx,\vecxi,t)$. A 
direct calculation shows that (\ref{trans1}) can be expressed as
\begin{equation}
\label{trans1a}
\vecnab_{\vecp}H^+(\vecnab_{\vecx}S^+,\vecx)\vecnab_{\vecx} +
\frac{\partial}{\partial t} + \frac{1}{2}\vecnab_{\vecx} \left[ 
\vecnab_{\vecp}H^+(\vecnab_{\vecx}S^+,\vecx) \right] 
+\ui M_+(\vecnab_{\vecx}S^+,\vecx) \ ,
\end{equation}
where we defined the traceless hermitian $2\times 2$ matrix $M_+$ as
\begin{equation}
\label{M+def}
M_+ (\vecp,\vecx) := -\frac{ec}{2\ve(\vecp,\vecx)} \left[ \vecB(\vecx)
                     +\frac{c}{\ve(\vecp,\vecx)+mc^2} \left( \vecE (\vecx)
                     \times \left(\vecp -\frac{e}{c}\vecA(\vecx) \right)
                     \right) \right] \vecsig \ . 
\end{equation}
We remark that since $M_+$ is a linear combination of the Pauli matrices,
$\ui M_+$ is an element of the Lie algebra su(2). In order to emphasize 
this point, and for later purposes, we introduce the following notation,
\begin{equation}
\label{Rdef}
M_+ (\vecp,\vecx) =: \frac{1}{2}\,\vecR(\vecp,\vecx)\,\vecsig \ .
\end{equation}

According to the usual convention, the resulting equation for $\widetilde V$
is called (first order) transport equation. At this point a comparison
with the case of the Schr\"odinger equation seems to be instructive, 
see e.~g.\ \cite{Rob87}. There the lowest order amplitude $a_0^{Schr.}$ 
is a scalar function that is required to be a solution of a transport 
equation which arises upon acting on $a_0^{Schr.}$ with the equivalent 
to (\ref{trans1a}), but where $M_+\equiv 0$. The solution, fixed by the 
initial condition $a_0^{Schr.}=1$ at $t=0$, is well known to be  
\begin{equation}
\label{a0Sch}
a_0^{Schr.}(\vecx,\vecy,t;\vecxi) = \sqrt{\det\left(\frac{\partial^2  S}
{\partial x_k\partial\xi_l}(\vecx,\vecxi,t)\right)} \ .
\end{equation}
Returning to the present situation described by (\ref{trans1a}), the 
following ansatz for the solution of the transport equation therefore 
seems to be appropriate,
\begin{equation}
\label{D+def}
\widetilde V (\vecx,\vecxi,t) = \sqrt{\det\left(\frac{\partial^2
S^+}{\partial x_k\partial\xi_l}(\vecx,\vecxi,t)\right)}\
U_+(\vecx,\vecxi,t) \ .
\end{equation}
As a consequence, the transport equation for $\widetilde V$ implies that 
the matrix-valued function $U_+$ has to solve the equation
\begin{equation}
\label{transVtilde}
\left[ \vecnab_{\vecp}H^+(\vecnab_{\vecx}S^+,\vecx)\vecnab_{\vecx} +
\frac{\partial}{\partial t} + \ui M_+(\vecnab_{\vecx}S^+,\vecx) \right]
\,U_+(\vecx,\vecxi,t) = 0 \ ,
\end{equation}
with initial condition $U_+(\vecx,\vecxi,0)=V^\dagger (\vecxi,\vecx)$.

It is not necessary to solve (\ref{transVtilde}) in full generality because
the amplitude $a_0^+$ enters the semiclassical expression for the time
evolution only at stationary points $\vecxi_j$ of the phase $\phi^+$.
We thus first employ the method of stationary phase to the oscillatory
integral (\ref{oscillint}), in which we only take the lowest order 
contributions $a_0^\pm$ to the amplitudes into account.  According to 
(\ref{genfct})--(\ref{generfct}), the stationary points of the phase, 
i.~e., the solutions $\vecxi_j$ of
\begin{equation}
\label{statjdef}
\vecnab_{\vecxi}\phi^\pm (\vecx,\vecy,t;\vecxi_j) = \vecnab_{\vecxi}
S^\pm(\vecx,\vecxi_j,t) -\vecy = 0\ ,
\end{equation}
determine solutions $(\vecP(t';\vecxi_j,\vecy),\vecX(t';\vecxi_j,\vecy))$,
$0\leq t'\leq t$, of the classical equations of motion from 
$(\vecxi_j,\vecy)$ to $(\vecp,\vecx)$ in time $t$. Their projections 
$\vecX(t';\vecxi_j,\vecy)$ to configuration space will also be denoted as 
$\ga_{xy}^\pm$. At a stationary point the phase is given by Hamilton's 
principal function for the trajectory $\ga_{xy}^\pm$ corresponding to 
$\vecxi_j$,
\begin{equation}
\label{Hamprifun}
\phi^\pm (\vecx,\vecy,t;\vecxi_j) = S^\pm(\vecx,\vecxi_j,t) -\vecy\vecxi_j
= \int_0^t L^\pm \left( \vecX(t'),\dot{\vecX}(t')\right) \ud t' 
=: R^\pm_{\ga_{xy}^\pm}(\vecx,\vecy,t) \ ,
\end{equation}
where $L^\pm$ is the Lagrangian associated with $H^\pm$. 

The method of stationary phase requires the amplitudes $a_0^\pm$, and 
hence $U_\pm$, evaluated at the stationary points $\vecxi_j$. In this 
situation $\vecx$ has to be considered as the end point of the corresponding
trajectory $\ga^+_{xy}$, i.~e., $\vecx=\vecX(t)$. In (\ref{transVtilde}) 
the first two terms can therefore be understood as a total time derivative 
along $\vecX(t)$,
\begin{equation}
\label{trajecdiff}
\vecnab_{\vecp}H^+ (\vecP(t),\vecX(t))\vecnab_{\vecx} + \frac{\partial}
{\partial t} = \dot{\vecX}(t)\vecnab_{\vecx} + \frac{\partial}{\partial t}
=: \frac{\ud}{\ud t} \ .
\end{equation}
In order to solve (\ref{transVtilde}) at a stationary point $\vecxi_j$ we 
now introduce the ansatz
\begin{equation}
\label{d+def}
U_+ (\vecX(t),\vecxi_j,t) = 
d_+ (\vecP(t),\vecX(t))\,V^\dagger (\vecxi_j,\vecy) \ ,
\end{equation}
which immediately implies the initial condition 
\begin{equation}\label{d+init}
 d_+(\vecP(0),\vecX(0)) =\eins_2
\end{equation}
for the $2\times 2$ matrix $d_+$. Restricting (\ref{transVtilde}) 
to the trajectory $\vecX(t)$ and using the abbreviation (\ref{trajecdiff}) 
then shows that this matrix is required to solve 
\begin{equation}
\label{dtranseq}
\left[ \frac{\ud}{\ud t}+\ui M_+ (\vecP(t),\vecX(t))\right] \,
d_+ (\vecP(t),\vecX(t)) = 0 
\end{equation}
along the trajectory. In the next section we will demonstrate that $d_+$
can be interpreted as a semiclassical propagator for the spin degrees of
freedom. Occasionally, we will thus refer to (\ref{dtranseq}) as the
spin transport equation. A formal solution of this equation can be given in 
terms of a time-ordered exponential,
\begin{equation}
d_+ (t) = {\text T}\,\exp\left\{ -\ui\int_0^t M_+ (t')\ \ud t'\right\} \ .
\end{equation}
Since according to (\ref{M+def}) $\ui M_+$ takes values in the Lie algebra
su(2), the solution $d_+$ is an element of the group SU(2). Together with 
its connection to a classical spin, a further geometric interpretation of 
$d_+$ will be provided in the next section.

Combining (\ref{a0pm}), (\ref{D+def}), and (\ref{d+def}) finally yields 
the lowest-order amplitude at a stationary point as
\begin{equation}
\label{a0result}
\begin{split}
a_0^+ (\vecx,\vecy,t;\vecxi_j) 
= &\sqrt{\det\left(\frac{\partial^2 S^+}{\partial x_k\partial\xi_l}
   (\vecx,\vecxi_j,t)\right)} \\
  &V\bigl( \vecnab_{\vecx}S^+ (\vecx,\vecxi_j,t),\vecx \bigr)\,
   d_+\bigl(\vecnab_{\vecx}S^+ (\vecx,\vecxi_j,t),\vecx\bigr)\,
   V^\dagger (\vecxi_j,\vecy) \ .
\end{split}
\end{equation}
This expression has an obvious analogue for the index $-$, for which the 
$2 \times 2$ matrix $M_-$ is given by 
\begin{equation}
M_- (\vecp,\vecx) := \frac{ec}{2\ve(\vecp,\vecx)} \left[ \vecB(\vecx)
                     -\frac{c}{\ve(\vecp,\vecx)+mc^2} \left( \vecE (\vecx)
                     \times \left(\vecp -\frac{e}{c}\vecA(\vecx) \right)
                     \right) \right] \vecsig \ .   
\end{equation}
Comparing (\ref{a0result}) with 
(\ref{a0Sch}) reveals that, apart from the occurrence of two classical
Hamiltonians, the only difference to the case of the semiclassical 
propagator for the Schr\"odinger equation is given by the appearence of 
the last three factors on the r.h.s.\ of (\ref{a0result}), which we  
also abbreviate as $V_t\,d_+ V_0^\dagger$. With the further short-hand
\begin{equation}
D_{\ga^\pm_{xy}} := \left|\det\left(-\frac{\partial^2 R^\pm_{\ga^\pm_{xy}}}
{\partial x_k\partial y_l}(\vecx,\vecy,t)\right)\right|^{1/2}
\end{equation}
the result of the method of stationary phase applied to (\ref{oscillint})
finally reads (for $t\neq 0$)
\begin{equation}
\label{Ksemicl}
\begin{split}
K(\vecx,\vecy,t) = \frac{1}{(2\pi\ui\hbar)^{3/2}}
                   &\left\{\sum_{\ga_{xy}^+} V_t\,d_+ V_0^\dagger\,
                    D_{\ga^+_{xy}}\,\ue^{\frac{\ui}{\hbar}R^+_{\ga^+_{xy}}
                    -\ui\frac{\pi}{2}\nu_{\ga^+_{xy}}}\, [1+O(\hbar)]
                    \right. \\
                   &\left.\quad +\sum_{\ga_{xy}^-} W_t\,d_- W_0^\dagger\,
                    D_{\ga^-_{xy}}\,\ue^{\frac{\ui}{\hbar}R^-_{\ga^-_{xy}}
                    -\ui\frac{\pi}{2}\nu_{\ga^-_{xy}}}\, [1+O(\hbar)] 
                    \right\} \ ,                                
\end{split}
\end{equation}
where, as in the case of the Schr\"odinger equation, $\nu_{\ga^\pm_{xy}}$
denotes the Morse index of the trajectory $\ga^\pm_{xy}$. Notice that 
the two sums extend over the classical trajectories of relativistic
point particles. The spin degrees of freedom, which manifest themselves
in the Dirac equation through the matrix character of the Hamiltonian, 
enter in leading semiclassical order only through the factors 
$V_t\,d_+ V_0^\dagger$ and $W_t\,d_- W_0^\dagger$, respectively. Since 
these contain no $\hbar$, a classical interpretation of them seems to
be in order. In the next section we will indeed obtain from (\ref{dtranseq})
an evolution equation for a classical spin which is transported through
the external fields along the particle trajectories $\ga^\pm_{xy}$.

As our final step in this section, we now modify (\ref{Ksemicl}) to yield
a semiclassical representation for the truncated time evolution kernel
$K_\chi$. According to (\ref{Kchismcl}), to this end one must apply the
$4\times 4$ matrix $\chi(H_D(\vecnab_{\vecx}S^\pm,\vecx))$ to the 
ampliude $a_0^\pm (\vecx,\vecy,t;\xi)$. Since the spectral representation
$H_D =H^+ P_+ +H^- P_-$ of the symbol matrix implies
\begin{equation}
\chi(H_D) = \chi(H^+)\,P_+ + \chi(H_-)\,P_- 
= \chi(H^+)\,V\,V^\dagger + \chi(H^-)\,W\,W^\dagger\ ,
\end{equation}
representing the amplitudes as in (\ref{a0pm}) immediately shows that
\begin{equation}
\chi(H_D)\,a_0^\pm = \chi( H^\pm)\,a_0^\pm \ .
\end{equation}
Since therefore the matrix-valued amplitudes are only multiplied by a
scalar factor, which is moreover constant along any classical trajectory
$\ga_{xy}^\pm$, the only modification of (\ref{Ksemicl}) consists of
an inclusion of the factors $\chi(E_{\ga_{xy}^\pm})$ under the sums over 
the trajectories, where $E_{\ga_{xy}^\pm}=H^\pm(\vecP(t),\vecX(t))$ 
denotes the (constant) energy of $\ga_{xy}^\pm$.

%%%%%%%%%%%%%% section 4 %%%%%%%%%%%%%%%

\section{Geometry of semiclassical spin transport}

\label{sec4}

The semiclassical representation for the time evolution kernel that
was derived in the previous section is principally determined by the
classical dynamics of relativistic point particles. This is true to
the extent that it essentially suffices to solve the equations of
motion generated by the scalar Hamiltonians $H^\pm$. Indeed, the 
semiclassically dominating $\hbar$-dependent phases occurring in 
(\ref{Ksemicl}) are completely fixed by Hamilton's principal functions 
$R^\pm$ of the translational motion. The latter decouples from the spin 
dynamics that, moreover, only contributes to the $\hbar$-independent 
amplitudes in (\ref{Ksemicl}).

In this section we will investigate the dynamics of the spin degrees of
freedom more closely. In particluar, we will interpret the SU(2)
matrix $d_+$ as a semiclassical time evolution operator for spin and, 
furthermore, relate it to a classical spin evolving according to a 
classical equation of motion. In order to prepare for this, we first have 
to clarify the geometrical setting of spin transport along classical 
particle trajectories further. To this end we consider the $H^+$-eigenbundle 
$E^+$ over phase space. It is defined as the disjoint collection of the 
eigenspaces 
\begin{equation}
E^+_{(\vecp,\vecx)}:=P_+(\vecp,\vecx)\,\kz^4
\end{equation}
of the symbol matrix $H_D(\vecp,\vecx)$ corresponding to the eigenvalue 
$H^+(\vecp,\vecx)$ for each point $(\vecp,\vecx)$ in phase space. Since
$H_D$ is hermitian, the decomposition of $\kz^4$ into the $H^+$- and
$H^-$-eigenspaces is orthogonal with respect to the canonical scalar 
product. This can hence be projected to the fibres $E^+_{(\vecp,\vecx)}$, 
so that $E^+$ is a $\kz^2$ vector bundle with structure group U(2). 
In terms of the orthonormal basis $\{e_1(\vecp,\vecx),e_2(\vecp,\vecx)\}$ 
used to define the matrix $V(\vecp,\vecx)$ in (\ref{VWdef}), a section 
in the eigenbundle can be represented as
\begin{equation}
v(\vecp,\vecx) = \sum_{k=1}^2 v_k(\vecp,\vecx)\,e_k(\vecp,\vecx) = 
V(\vecp,\vecx) \begin{pmatrix} v_1(\vecp,\vecx) \\ v_2(\vecp,\vecx)
\end{pmatrix} \ . 
\end{equation}

To lowest semiclassical order the spin dynamics is governed by the transport
equation defined with the differential operator (\ref{trans1}). This requires
sections in the eigenbundle $E^+$ that are lifts of trajectories in phase 
space. We therefore now consider the following differentiation on sections $v$,
which we evaluate at $(\vecp,\vecx)=(\vecnab_{\vecx}S^+
(\vecx,\vecxi,t),\vecx)$. Then
\begin{eqnarray}
\tilde v(\vecnab_{\vecx}S^+,\vecx)
&:=&P_+(\vecnab_{\vecx}S^+,\vecx)\,\left[ c\vecalph\vecnab_{\vecx}+
 \frac{\partial}{\partial t} \right]\,v(\vecnab_{\vecx}S^+,\vecx) \\
&=&V(\vecnab_{\vecx}S^+,\vecx)\,V^\dagger(\vecnab_{\vecx}S^+,\vecx)
 \left[ c\vecalph\vecnab_{\vecx}+\frac{\partial}{\partial t} \right]
 V(\vecnab_{\vecx}S^+,\vecx)
 \begin{pmatrix} v_1(\vecnab_{\vecx}S^+,\vecx) \\ 
 v_2(\vecnab_{\vecx}S^+,\vecx)\end{pmatrix}  \nonumber
\end{eqnarray}
is again a section in $E^+$ in the above sense. According to
(\ref{trans1}) and (\ref{trans1a}) an expansion of $\tilde v$ in the 
gliding basis $\{e_1,e_2\}$ yields the coefficients
\begin{equation}
\begin{split}
\begin{pmatrix} \tilde v_1(\vecnab_{\vecx}S^+,\vecx) \\ 
\tilde v_2(\vecnab_{\vecx}S^+,\vecx) \end{pmatrix} = 
&\biggl( \vecnab_{\vecp}H^+(\vecnab_{\vecx}S^+,\vecx)\vecnab_{\vecx} 
 +\frac{\partial}{\partial t} + \frac{1}{2}\vecnab_{\vecx} \left[ 
 \vecnab_{\vecp}H^+(\vecnab_{\vecx}S^+,\vecx) \right]  \\
&\quad +\ui M_+(\vecnab_{\vecx}S^+,\vecx) \biggr)
 \begin{pmatrix} v_1(\vecnab_{\vecx}S^+,\vecx) \\ 
 v_2(\vecnab_{\vecx}S^+,\vecx)\end{pmatrix} \ .
\end{split}
\end{equation}
Furthermore, a separation analogous to (\ref{D+def}),
\begin{equation}
v(\vecnab_{\vecx}S^+,\vecx) = \sqrt{\det\left(\frac{\partial^2S^+}
{\partial x_k\partial\xi_l}(\vecx,\vecxi,t)\right)}\,
u(\vecnab_{\vecx}S^+,\vecx) \ ,
\end{equation}
leads via (\ref{transVtilde}) to the definition of a covariant 
differentiation $\cD^+$ on the section $u$,
\begin{equation}
\label{cDdef}
\begin{split}
\cD^+ &u(\vecnab_{\vecx}S^+,\vecx) := \\ 
&V(\vecnab_{\vecx}S^+,\vecx) \left( \vecnab_{\vecp}H^+(\vecnab_{\vecx}
 S^+,\vecx)\vecnab_{\vecx} + \frac{\partial}{\partial t} + 
 \ui M_+(\vecnab_{\vecx}S^+,\vecx) \right) 
 \begin{pmatrix} u_1(\vecnab_{\vecx}S^+,\vecx) \\ 
 u_2(\vecnab_{\vecx}S^+,\vecx)\end{pmatrix} \ .
\end{split}
\end{equation}
If one now restricts the variable $\vecxi$ to the stationary points 
$\vecxi_j$ arising in (\ref{statjdef}), the first two terms in (\ref{cDdef}) 
again yield a differentiation along the trajectory $\ga_{xy}^+$ associated 
with $\vecxi_j$, compare (\ref{dtranseq}). One then also recognizes
$\ui M_+$ as the su(2)-valued connection coefficient arising for the
connection associated with $\cD^+$. A section $u$ given along a solution 
of the classical equations of motion is therefore parallel, if it is a 
solution of
\begin{equation}
\cD^+ u(\vecP(t),\vecX(t)) = 0 \quad\text{with}\quad 
u_k(\vecP(0),\vecX(0)) = u_{k,0} \ .
\end{equation}
According to (\ref{dtranseq}) such a solution can also be represented as
\begin{equation}
\label{dholon}
\begin{pmatrix} u_1(\vecP(t),\vecX(t)) \\ u_2(\vecP(t),\vecX(t))\end{pmatrix}
= d_+ (\vecP(t),\vecX(t))\,\begin{pmatrix} u_{1,0} \\ u_{2,0}\end{pmatrix}\ .
\end{equation}
Thus, geometrically $d_+\in\text{SU(2)}$ describes the parallel transport 
in $E^+$ defined by the connection arising from $\cD^+$. Along periodic 
orbits in phase space, $d_+$ therefore yields the holonomy of this 
connection. This geometric interpretation ensures that the combination 
$V_t\,d_+\,V_0^\dagger$, which appears in the semiclassical expression 
(\ref{Ksemicl}) for the time evolution kernel, is invariant under unitary 
base changes in $E^+$. In physical terms, $d_+$ can be interpreted as the 
semiclassical time evolution operator for two-spinors in the representation 
defined via the gliding basis $\{e_1,e_2\}$. The unitarity of $d_+$ then 
implies that the norm $|u_{1,0}|^2 +|u_{2,0}|^2$ of the initial two-spinor 
$(u_{1,0},u_{2,0})^T\in\kz^2$ is preserved under this evolution. In the 
following we will therefore always consider normalised sections $u$ in $E^+$.  

We would now like to compare the above construction with the connections
that appear in the analysis of Littlejohn and Flynn \cite{LitFly91b}, and
Emmrich and Weinstein \cite{EmmWei96}. Since the eigenbundle $E^+$ is a 
(non-trivial) subbundle of the trivial $\kz^4$ bundle over phase space, a 
natural connection in $E^+$ arises by projecting the trivial covariant 
differentiation of the $\kz^4$ bundle to $E^+$. Along a trajectory 
$(\vecP(t),\vecX(t))$ this construction reads
\begin{equation}
P_+ (\vecP(t),\vecX(t))\left[\frac{\ud}{\ud t}u(\vecP(t),\vecX(t))\right]
=: \cD^B u(\vecP(t),\vecX(t))\ ,
\end{equation}
when applied to a section $u$ in $E^+$. In analogy to (\ref{cDdef}) one 
then obtains a covariant differentiation in terms of the coefficients
$u_k$ with respect to the gliding basis $\{e_1,e_2\}$. The connection 
coefficient $\ui M_B$ that replaces $\ui M_+$ can be calculated as
\begin{equation}
\begin{split}
M_B(\vecp,\vecx)=
 &\frac{ec^2}{2\,\ve(\vecp,\vecx)[\ve(\vecp,\vecx)+mc^2]} \left[ 
  (\vecp -\tfrac{e}{c}\vecA(\vecx))\times \vecE(\vecx) \right] \vecsig \\
 &+\frac{ec^3}{2\,\ve^2(\vecp,\vecx)[\ve(\vecp,\vecx)+mc^2]} \left[
  (\vecp -\tfrac{e}{c}\vecA(\vecx)) \times 
  \bigl( (\vecp-\tfrac{e}{c}\vecA(\vecx))\times\vecB(\vecx) \bigr) \right] 
  \vecsig \ .
\end{split}
\end{equation}
That way one defines a connection on $E^+$ that bears some similarities 
to the adiabatic connection in quantum mechanics. The latter has been 
identified by Simon \cite{Sim83} to produce the Berry phase \cite{Ber84}, 
see also \cite{EmmWei96}. On this ground, Littlejohn and Flynn 
\cite{LitFly91a,LitFly91b} introduced the notion of a Berry term for the 
analogue to our $M_B$ in the case of principal symbol matrices with no 
globally degenerate eigenvalues. In the present context, the twofold 
degenerate eigenvalue $H^+$ forces $M_B$ to take values in su(2), which 
hence leads to a connection with SU(2) holonomy. We emphasize, however, 
that no adiabatic approximation is made after one has arrived at the
transport equations (\ref{orderk}) and (\ref{transVtilde}), respectively.
Nevertheless, following \cite{LitFly91a,LitFly91b} in spirit, we refer 
to $M_B$ as the SU(2)-Berry term, although regarding adiabatic 
approximations this notation is sligthly misleading. 

The difference between the two covariant differentiations $\cD^+$ and 
$\cD^B$ can be expressed in terms of a connection coefficient $M_C =
M_+ -M_B$. This expression, which Littlejohn and Flynn \cite{LitFly91b} 
refer to as the no-name term, has been identified by Emmrich and Weinstein 
\cite{EmmWei96} to be related to the curvature associated with $\cD^B$.
They moreover showed that
\begin{equation}
\label{MCdef}
M_C (\vecp,\vecx) = -\frac{\ui}{2}V^\dagger(\vecp,\vecx)\, 
\{P_+,H_D -H^+\eins_4\}(\vecp,\vecx)\ V(\vecp,\vecx) \ ,
\end{equation}
where the Poisson bracket for two matrix-valued functions $A,B$ on phase
space is defined as
\begin{equation}
\{A,B\}(\vecp,\vecx) := \vecnab_{\vecp}A(\vecp,\vecx)\,\vecnab_{\vecx}
B(\vecp,\vecx)-\vecnab_{\vecp}B(\vecp,\vecx)\,\vecnab_{\vecx}A(\vecp,\vecx)\ .
\end{equation}
Notice that the ordering of the matrices is important. Upon explicitly
calculating the r.h.s.\ of (\ref{MCdef}) one can verify that $M_+ =M_B
+M_C$.

We now define a time-dependent spin operator $\vecSig (t)$, whose
components act on $\kz^2$, as
\begin{equation}
\label{defSig}
\vecSig (t) := d_+^\dagger (\vecP(t),\vecX(t))\,\vecsig\,
               d_+ (\vecP(t),\vecX(t)) \ ,
\end{equation}
so that $\vecSig(0)=\vecsig$. According to (\ref{dtranseq}), the dynamics
of the spin operator $\vecSig (t)$ is governed by 
\begin{equation}
\frac{\ud}{\ud t}\vecSig (t) = 
\ui\,d_+^\dagger(t) \left[ M_+(t)\,\vecsig - \vecsig\,M_+(t) \right]
d_+ (t)\ .
\end{equation}
Introducing $\vecR(\vecp,\vecx)$ as in (\ref{Rdef}), the evolution equation 
for $\vecSig$ can be brought into the convenient form
\begin{equation}
\label{Sigevolve}
\frac{\ud}{\ud t}\vecSig (t) = \vecR(t) \times \vecSig (t) \ ,
\end{equation}
which describes the precession of the vector $\vecSig$ about the 
instantaneous axis defined by $\vecR$. In order to obtain an object that 
can be considered as a classical spin, one introduces the expectation 
value of $\vecSig$ in a two-spinor state prescribed by 
$(u_{1,0},u_{2,0})^T\in\kz^2$,
\begin{equation}
\label{spinex}
\vecs(t)  := (\overline{u}_{1,0},\overline{u}_{2,0})\,\vecSig (t)
             \begin{pmatrix} u_{1,0} \\ u_{2,0} \end{pmatrix} 
           = (\overline{u}_1(t),\overline{u}_2(t))\,\vecsig
             \begin{pmatrix} u_1(t) \\ u_2(t) \end{pmatrix}   
           = \begin{pmatrix} 2\,\re (\overline{u}_1(t)\,u_2(t)) \\
             2\,\im (\overline{u}_1(t)\,u_2(t)) \\
             |u_1(t)|^2 - |u_2(t)|^2 \end{pmatrix} \ ,
\end{equation}
which obviously also solves (\ref{Sigevolve}), i.~e.,
\begin{equation}
\label{BMT}
\frac{\ud}{\ud t}\vecs = \vecs \times 
\left( \frac{ec}{\ve}\vecB - \frac{ec^2}{\ve(\ve +mc^2)}\left(\vecp -
\frac{e}{c}\vecA\right)\times \vecE \right) \ , 
\end{equation}
where all quantities have to be taken along a given trajectory $\ga^+_{xy}$.
This classical equation is well known to describe the precession of a 
spinning particle in external electromagnetic fields $\vecE$ and $\vecB$, as 
has already been demonstrated by Thomas \cite{Tho27}. In \cite{BarMicTel59} it 
is rederived in a manifestly covariant formulation, and therefore also goes 
under the notion of BMT equation. In the present context, (\ref{BMT}) can
be viewed as a manifestation of the Ehrenfest theorem for spin because
here $\vecs$ stands for any expectation value of the spin operator $\vecSig$.
Notice that since we deliberately omitted a factor of $\hbar/2$ in the 
definition (\ref{defSig}) of the spin operator, the classical spin
vector $\vecs$ is of (constant) unit length. 

So far we have provided geometrical and physical interpretations of $d_+$.
Yet, for the purpose of the semiclassical representation (\ref{Ksemicl}) 
of the time evolution kernel one needs to calculate the SU(2) matrix 
\begin{equation}
d_+ (\vecP(t),\vecX(t)) =: 
\begin{pmatrix} \al(t) & -\overline{\be}(t) \\ 
\be(t) & \overline{\al}(t) \end{pmatrix}\ , \qquad
|\al(t)|^2 + |\be(t)|^2 =1 \ ,
\end{equation}
for each classical trajectory $\ga^+_{xy}$. This can be achieved along
the lines presented above, if one chooses the initial condition 
$u(\vecxi_j,\vecy)=e_1(\vecxi_j,\vecy)$ for a section $u$ solving 
$\cD^+ u=0$. In the two-spinor representation for $u$ this initial condition 
reads $(u_{1,0},u_{2,0})=(1,0)$, and thus (\ref{dholon}) implies that 
\begin{equation}
\label{albeid}
\begin{pmatrix} \al(t) \\ \be(t) \end{pmatrix} = 
\begin{pmatrix} u_1(\vecP(t),\vecX(t)) \\ u_2(\vecP(t),\vecX(t))\end{pmatrix}
\ .
\end{equation}
The spin expectation value $\vecs(t)$ corresponding to this choice then
follows from (\ref{spinex}) as
\begin{equation}
\label{spinexp}
\vecs(t) = (\overline{\al}(t),\overline{\be}(t))\,\vecsig
           \begin{pmatrix} \al(t) \\ \be(t) \end{pmatrix} =
           \begin{pmatrix} 2\,\re (\overline{\al}(t)\,\be(t)) \\
           2\,\im (\overline{\al}(t)\,\be(t))  \\
           |\al(t)|^2 - |\be(t)|^2 \end{pmatrix}\ ,\quad\text{with}\quad
\vecs(0) = \begin{pmatrix} 0 \\ 0 \\ 1 \end{pmatrix}\ .
\end{equation}
Mathematically, this physically motivated construction can be viewed as a 
map from $d_+\in\text{SU(2)}$ to $\vecs\in {\rm S^2}\subset\rz^3$. Indeed, it 
is known as the Hopf map $\pi_H :\,\text{SU(2)}\rto {\rm S^2}$, which yields a 
U(1) principal fibre bundle over ${\rm S^2}$. 

In order to calculate $d_+(t)$ for a given trajectory $\ga_{xy}^+$ one
first has to integrate the evolution equation (\ref{BMT}) for the precession 
of a classical spin along $\ga_{xy}^+$ with initial condition $\vecs(0)=
(0,0,1)^T$. According to the relation $\vecs(t)=\pi_H (d_+(t))$ one thus 
has determined two of the three real degrees of freedom of $d_+(t)$ by 
classical means. The third degree of freedom can only be 
reconstructed, if one is able to lift the curve $\vecs(t)$ in ${\rm S^2}$ to 
SU(2). To this end one requires a connection on the U(1) bundle over 
${\rm S^2}$ that is provided by the Hopf map. The two degrees of freedom of the
classical spin $\vecs$ can be related to convenient coordinates on ${\rm S^2}$ 
once one notices 
that the normalisation forces $|\al|$ to range in the interval $[0,1]$. 
Therefore, a variable $\theta\in [0,\pi]$ can be introduced such that 
\begin{equation}
\label{thetadef}
|\al| = \cos\tfrac{\theta}{2} \qquad\text{and}\qquad 
|\be| = \sin\tfrac{\theta}{2} \ . 
\end{equation}
Together with the phases of $\al=|\al|\,\ue^{\ui\eta}$ and $\be=|\be|\,
\ue^{\ui\la}$ one thus has three degrees of freedom at hand to describe $d_+$. 
A representation of the classical spin $\vecs$ in terms of the
variables $(\theta,\eta,\la)$ then follows from (\ref{spinexp}),
\begin{equation}
\label{spheric}
\vecs = \begin{pmatrix} \sin\theta\,\cos\phi \\ \sin\theta\sin\phi \\
                        \cos\theta \end{pmatrix} 
\ , \quad\text{with}\quad \phi:=\la -\eta \ ,
\end{equation}
so that $(\theta,\phi)$ can be identified as the usual spherical coordinates 
on ${\rm S^2}$. In principle, one could now choose $\la+\eta$ as the 
additional
spin degree of freedom discussed above. However, in order to identify the
connection on the Hopf fibration it turns out that a choice of either $\eta$ 
or $\la$ separately is more approppriate. Since the north pole of ${\rm S^2}$ 
corresponds to $(\al_N,\be_N)=(1,0)$ and the south pole to $(\al_S,\be_S)=
(0,1)$, the phase $\eta$ is ill-defined at the south pole, whereas $\la$ is 
ill-defined at the north pole. Following the procedure well known from the 
analysis of magnetic monopoles by Wu and Yang \cite{WuYan75}, we now choose 
$\eta$ as a fibre coordinate for base points on the northern hemisphere 
$U_N\subset {\rm S^2}$ and, correspondingly, $\la$ for base points on the 
southern hemisphere $U_S\subset {\rm S^2}$. Since the connection we want to 
identify is fixed by the 
requirement that the lifted curve on SU(2) be the solution $d_+(t)$ of 
(\ref{dtranseq}), an explicit expression for the connection coefficients 
can be derived from the equation
\begin{equation}
\frac{\ud}{\ud t}\begin{pmatrix} \al(t) \\ \be(t) \end{pmatrix} =
-\ui M_+ (t)\begin{pmatrix} \al(t) \\ \be(t) \end{pmatrix}
\end{equation}
that follows from the identification (\ref{albeid}). Multiplying this with
$(\overline{\al}(t),\overline{\be}(t))$ from the left now yields 
\begin{equation}
|\al(t)|^2 \frac{\ud}{\ud t}\eta(t) 
+ \bigl(1-|\al(t)|^2\bigr)\,\frac{\ud}{\ud t} \la(t) 
= -\frac{1}{2}\,\vecR(t)\vecs(t) \ .
\end{equation}
Exploiting (\ref{thetadef}) and (\ref{spheric}) then results in equations 
for $\eta$ on $U_N$ and $\la$ on $U_S$, respectively,
\begin{equation}
\begin{align}
\label{etadgl}
\frac{\ud}{\ud t}\eta(t) &= -\frac{1}{2}\,\vecR(t)\vecs(t) - \frac{1}{2}\,
                            \bigl(1-\cos\theta(t)\bigr)\,\frac{\ud}{\ud t}
                            \phi(t) \ , \\
\label{lambdadgl}
\frac{\ud}{\ud t}\la(t)  &= -\frac{1}{2}\,\vecR(t)\vecs(t) + \frac{1}{2}\,
                            \bigl(1+\cos\theta(t)\bigr)\,\frac{\ud}{\ud t}
                            \phi(t) \ .
\end{align}
\end{equation}
According to the initial condition in (\ref{spinexp}) the motion of the 
classical spin $\vecs$ starts at the north pole so that for sufficiently 
small times $\vecs(t)\in U_N$. Thus the phase $\eta$ should be used, whose 
initial condition $\eta(0)=0$ follows from (\ref{d+init}). This allows an 
immediate integration of (\ref{etadgl}),
\begin{equation}
\label{etaoft}
 \eta(t) = -\frac{1}{2} \int_0^t \vecR(t') \vecs(t') \ \ud t'
           -\frac{1}{2} \int_0^t \big( 1 - \cos\theta(t') \big) 
            \,\frac{\ud\phi}{\ud t'}(t') \ \ud t' \ .
\end{equation}
The first term on the r.h.s.\ of (\ref{etaoft}) is a dynamical phase 
associated with the classical energy of a magnetic moment in given 
electromagnetic fields, whereas the second term%
\footnote[1]{We remark that in contrast to equation (21) of our recent 
             Letter \cite{BolKep98} the sign of this term should be as 
             in (\ref{etaoft}).} %
is a geometric phase. The
latter can be further characterised once one takes into account that 
$\eta$ is ill-defined at the south pole so that the phase $\lambda$ should 
be used instead of $\eta$ as soon as $\vecs$ 
enters $U_S$, say, at a time $t_0$. One then has to integrate 
(\ref{lambdadgl}) with initial condition $\lambda(t_0) = \phi(t_0) + 
\eta(t_0)$. To this end the one-form $-\frac{1}{2}(1-\cos\theta)\ud\phi$, 
which constitutes the geometric part of (\ref{etaoft}), has to be 
replaced by $\frac{1}{2}(1+\cos\theta)\ud\phi$, see (\ref{lambdadgl}). 
Since these two expressions are the gauge potentials of a magnetic monopole,
see \cite{WuYan75}, we can now formally identify the geometric part of 
(\ref{etaoft}) to be caused by a magnetic monopole of strength $-1/2$ 
situated at the origin of the sphere. We again emphasize that in the above 
consideration no adiabaticity assumption was made. Therefore, although the 
result for the geometric phase is strikingly similar to the Berry phase 
of a precessing quantum mechanical spin \cite{Ber84}, the geometric part 
of $\eta$ is rather of the more general type discussed by Aharonov and 
Anandan \cite{AhaAna87}.

Having integrated the equations for the classical spin and for the 
additional phase, we are now able to present the SU(2) matrix $d_+$ in 
the form 
\begin{equation}
\label{d+result}
 d_+(\vecP(t),\vecX(t)) = \left( \begin{matrix}
                 \cos \left( \tfrac{\theta}{2} \right) \ue^{\ui\eta} &
                -\sin \left( \tfrac{\theta}{2} \right) \ue^{-\ui(\eta+\phi)} \\
                 \sin \left( \tfrac{\theta}{2} \right) \ue^{\ui(\eta+\phi)} &
                 \cos \left( \tfrac{\theta}{2} \right) \ue^{-\ui\eta}
                          \end{matrix} \right) ,
\end{equation}
where $(\theta,\phi)$ are spherical coordinates for $\vecs(t)$, and $\eta$ 
is given by (\ref{etaoft}), if $\vecs(t) \in U_N$, and by $\eta(t) = 
\lambda(t) - \phi(t)$, if $\vecs(t) \in U_S$. With this explicit formula 
for $d_+$ all terms entering the semiclassical propagator (\ref{Ksemicl})
to leading order in $\hbar$ are completely defined in terms of classical 
quantities.

%%%%%%%%%%%%%% section 5 %%%%%%%%%%%%%

\section{Semiclassical trace formula}

\label{sec5}

After having obtained the fairly explicit expression (\ref{Ksemicl}) for
the time evolution kernel together with an interpretation in terms
of classical quantities, our ultimate goal now is to set up a 
semiclassical trace formula for the Dirac equation. In the case of the 
Schr\"odinger equation, Gutzwiller's original approach was to express the 
quantum mechanical density of states in terms of a sum over classical 
periodic orbits (or, more generally, over connected manifolds of periodic 
points of the classical flow). To this end he Fourier-transformed the 
semiclassical expression for the retarded time evolution kernel, which
he derived from its path integral representation, in order to obtain
a semiclassical approximation for the Green function. Subsequently he
performed the trace integral with the method of stationary phase. The
result then immediately yielded the semiclassical spectral density, see
\cite{Gut90} for details. Since that way one has to deal with several
singular objects, we here prefer to derive a regularised trace formula
that only takes finite quantities into account. The key relation for
this procedure is equation (\ref{pretrace}), in which we below use
a semiclassical representation for the truncated time evolution kernel
$K_\chi (\vecx ,\vecy ,t)$. We prefer this procedure, since first of all 
the energy localisation introduced through the truncation $\chi$ 
ensures that only the point spectrum of $\hat H_D$ contributes. Secondly, 
the test function $\tilde\varrho$ cuts off all periods of classical 
periodic orbits that exceed some maximal value $T_{max}$, since we request 
$\tilde\varrho$ to be compactly supported. This cut-off prohibits any 
possible clash of the two non-commuting asymptotics $\hbar\rto 0$ and 
$t\rto\infty$. Later $\tilde\varrho$ can be chosen to have an arbitrarily 
large, though compact, support. Possibly, this support condition can be
weakened by demanding a sufficiently strong decrease of $\tilde\varrho(t)$ as
$|t|\rto\infty$, see \cite{SieSte90b,Bol98}.

The semiclassical analysis presented at the end of section \ref{sec2}
already revealed that, as $\hbar\rto 0$, all contributions to 
(\ref{pretrace}) which exceed $O(\hbar^\infty)$ have to derive from
classical periodic orbits $\ga_p^\pm$. These are associated with the
stationary points $(\vecxi_{\ga_p^\pm},\vecx_{\ga_p^\pm},T_{\ga_p^\pm})$ 
of the phase $\phi^\pm +Et$ appearing in the integral (\ref{statphint}).
We also pointed out that the manifolds $M_0^\pm$ of stationary points 
$(\vecxi_0,\vecx_0,0)$ are related to the hypersurfaces $\Om_E^\pm$ of 
constant energy, which are composed of periodic points with trivial 
periods $T_0 =0$. We now assume the cleanness condition explained in
section \ref{sec2} and recall that in particular this implies that all 
further stationary points are such that the periods $T_{\ga_p^\pm}$ do not 
accumulate at $t =0$. It is then possible to choose a smooth function 
$h\in C_0^\infty (\rz)$ whose (connected) support contains the period 
$T_0 =0$, but no further period $T_{\ga_p^\pm}>0$. We furthermore require 
$h$ to fulfill $h(t)=1$ on some neighbourhood of $T_0 =0$. Upon introducing 
the partition of unity $1=h(t)+[1-h(t)]$ under the integral (\ref{pretrace}), 
the stationary points $(\vecxi_0,\vecx_0,0)$ are separated from the further 
stationary points $(\vecxi_{\ga_p^\pm},\vecx_{\ga_p^\pm},T_{\ga_p^\pm})$ 
associated with the non-trivial periodic orbits $\ga_p^\pm$. These two 
classes of stationary points will contribute to (\ref{pretrace}) in 
essentially different ways so that we will deal with them separately. Let 
our first concern hence be the calculation of the leading order contribution 
of the stationary points $(\vecxi_0,\vecx_0,0)$. We therefore consider 
the integral
\begin{equation}
\label{Bigsing}
\begin{split}
\frac{1}{2\pi (2\pi\hbar)^3}\int_{\rz^3}\int_{\rz^3}\int_{-\infty}^{+\infty}
h(t)\,\tilde\varrho(t)\,\mtr 
  & \left[ \chi\left( H_D (\vecnab_{\vecx}\phi^+,\vecx)\right)\,a_0^+\,
    \ue^{\frac{\ui}{\hbar}(\phi^+ +Et)}\right.  \\
  & \left. +\chi\left( H_D (\vecnab_{\vecx}\phi^-,\vecx)\right)\,a_0^-\,
    \ue^{\frac{\ui}{\hbar}(\phi^- +Et)} \right]\ \ud t\,\ud^3\xi\,\ud^3 x \ .
\end{split}
\end{equation}
Since the cut-off function $h$ ensures that no stationary points with 
$t\neq 0$ contribute, we expand the phases about $t=0$. Using the 
Hamilton-Jacobi equation for $S^\pm (\vecx,\vecxi,t)$, one finds
\begin{equation}
\label{phaseexp}
\begin{split}
\phi^\pm (\vecx,\vecx,t;\vecxi)+Et 
  &= S^\pm (\vecx,\vecxi,t) + Et - \vecx\vecxi \\
  &= t \left[ E+\frac{\partial S^\pm}{\partial t}(\vecx,\vecxi,0)\right] 
     + \frac{t^2}{2} \frac{\partial^2 S^\pm}{\partial t^2}
     (\vecx,\vecxi,0) + O(t^3) \\
  &= t \left[ E-H^\pm (\vecxi,\vecx) \right] + \frac{t^2}{2}
     \vecnab_{\vecp}H^\pm(\vecxi,\vecx)\,\vecnab_{\vecx}H^\pm(\vecxi,\vecx)
     + O(t^3) \ .
\end{split}
\end{equation}
At this stage we introduce in (\ref{Bigsing}) polar coordinates for
the variable $\vecxi$, i.~e., $\vecxi=\la\om$ with $\la:=|\vecxi|$ and
solid angle $\om\in {\rm S^2}\subset\rz^3$. This implies $d^3\xi =\la^2\,d\la\,
d\om$. We then employ the method of stationary phase to the integration 
over the variables $(t,\la)\in\rz\times\rz^+$. Stationary points are
therefore determined by a vanishing of the derivatives of (\ref{phaseexp})
with respect to $t$ and $\la$, respectively. Evaluated at $t=0$, the 
expression (\ref{phaseexp}) shows that this yields the condition 
$H^\pm (\la_0^\pm\om,x)=E$ to be fulfilled by the stationary points 
$(t_0^\pm =0,\la_0^\pm)$. Therefore, our first conclusion is that the 
integral over the remaining variables $(\vecx,\om)\in\rz^3\times {\rm S^2}$ is 
in fact restricted to the hypersurfaces $\Om_E^+$ and $\Om_E^-$, respectively. 
Carrying out the method of stationary phase further finally yields an 
asymptotic expansion of the integral (\ref{Bigsing}) as $\hbar\rto 0$ 
whose leading term can be determined explicitly in a straight forward 
manner,
\begin{equation}
\label{Weyl}
\chi(E)\,\frac{\tilde\varrho (0)}{2\pi}\,
\frac{2{\rm vol}\,(\Om_E^+) + {2\rm vol}
\,(\Om_E^-)}{(2\pi\hbar)^2}\,\left\{ 1+O(\hbar) \right\} \ ,
\end{equation}
where ${\rm vol}(\Omega_E^{\pm})$ denotes the volumes,
\begin{equation}
{\rm vol}(\Omega_E^{\pm}) = \int_{\rz^3} \int_{\rz^3} 
     \delta(H^{\pm}(\vecp,\vecx)-E) \ \ud^3p \, \ud^3x \ ,
\end{equation}
of the hypersurfaces $\Omega_E^{\pm}$.

For the computation of the contribution to (\ref{pretrace}) caused by the 
second class of stationary points, associated with the non-trivial classical 
periodic orbits, we proceed differently. As a starting point we consider 
the r.h.s.\ of (\ref{pretrace}), in which we introduce the semiclassical 
representation (\ref{Ksemicl}) of the truncated time evolution kernel. That 
is, we are going to evaluate
\begin{equation} 
\label{pocontrib}
\frac{1}{2\pi} \int_{\rz^3} \int_{-\infty}^{+\infty} [1-h(t)]\,
\tilde\varrho(t)\,\ue^{\frac{\ui}{\hbar}Et} \,\mtr K_\chi (\vecx,\vecx,t)
\ \ud t\,\ud^3 x
\end{equation}
with the method of stationary phase, after having inserted the r.h.s.\ 
of (\ref{Ksemicl}) for $K_\chi (\vecx,\vecx,t)$. Notice that the factor 
$1-h(t)$ cuts off a neighbourhood of $t=0$ and that the compactly supported 
function $\tilde\varrho$ further restricts the range of integration over $t$ 
to a bounded set. The semiclassical asymptotics (\ref{Ksemicl}), which was 
derived with $t$ held fixed, can therefore be used in (\ref{pocontrib}) 
without running into conflict with a need to perform $t\rto\infty$. Apart 
from the factors caused by the spin degrees of freedom in (\ref{Ksemicl}), 
and from the fact that the two relativistic classical Hamiltonians 
$H^\pm (\vecp,\vecx)$ determine the equations of motion for the translational 
degrees of freedom, the following calculation is closely parallel to the 
case of the Schr\"odinger equation \cite{Gut90}. 

As implied by the discussion at the end of section \ref{sec2}, the 
stationary points relevant for the semiclassical evaluation of 
(\ref{pocontrib}), which we now express as 
\begin{equation}
\label{Ksemiinsert}
\begin{split}
\frac{1}{2\pi(2\pi\ui\hbar)^{\frac{3}{2}}} 
   & \int_{\rz^3} \int_{-\infty}^{+\infty}
     [1-h(t)]\,\tilde\varrho(t)\, \left\{ \sum_{\ga_{xx}^+}\chi(E_{\ga_{xx}^+})
     \mtr \left( V_t\,d_+\,V_0^\dagger \right)\,D^+_{\ga_{xx}^+}\,
     \ue^{\frac{\ui}{\hbar}[R^+_{\ga_{xx}^+}+Et]-\ui\frac{\pi}{2}
     \nu_{\ga_{xx}}^+} \right. + \\
   & \left. \sum_{\ga_{xx}^-} \chi(E_{\ga_{xx}^-})
     \mtr \left( W_t\,d_-\,W_0^\dagger \right)\,
     D^-_{\ga_{xx}^-}\,\ue^{\frac{\ui}{\hbar}[R^-_{\ga_{xx}^-}+Et]-\ui
     \frac{\pi}{2}\nu_{\ga_{xx}}^-}\right\}\,\{1+O(\hbar)\}\ \ud t\,\ud^3 x\ ,
\end{split}
\end{equation}
derive from the non-trivial periodic orbits $\ga_p^\pm$. Indeed, the 
stationary points of the phases appearing in (\ref{Ksemiinsert}) are 
determined by 
\begin{equation}
\left[ \vecnab_{\vecx} R^\pm_{\ga_{xy}^\pm}(\vecx,\vecy,t) + \vecnab_{\vecy} 
R^\pm_{\ga_{xy}^\pm}(\vecx,\vecy,t) \right]_{\vecy=\vecx}=0 \quad \text{and}
\quad -\frac{\partial R^\pm_{\ga_{xx}^\pm}}{\partial t}(\vecx,\vecx,t)
=E \ .
\end{equation}
The first relation requires that initial and final momenta of those closed
trajectories $\ga_{xx}^\pm$ that contribute to the method of stationary phase
must coincide, so that these are indeed periodic. The second condition then 
picks out those periodic orbits that are contained in the hypersurfaces
$\Om_E^\pm$. At stationary points the phases then read
\begin{equation}
R^\pm_{\ga_p^\pm}(\vecx_{\ga_p^\pm},\vecx_{\ga_p^\pm},T_{\ga_p^\pm}) 
+ E\,T_{\ga_p^\pm} = 
\oint_{\ga_p^\pm}\vecp\ \ud\vecx =: S^\pm_{\ga_p^\pm} (E) \ .
\end{equation}
In order to apply the method of stationary phase to (\ref{Ksemiinsert})
we assume that all periodic orbits are isolated and non-degenerate. 
Strictly speaking, one now has to introduce a partition of unity that
separates the contributions of the isolated orbits. Since their periods
are known not to accumulate at some finite value $T$, this can be done
in the same manner as for the trivial period $T_0 =0$ above. 

The only difference to the case of the Schr\"odinger equation is caused
by the presence of the spin degrees of freedom. This is represented through 
the matrix trace under the integral in (\ref{Ksemiinsert}). At stationary 
points, however, $t=T_{\ga_p^\pm}$ and the periodicity implies that 
$V_{T_{\ga_p^+}}=V_0$. Thus a cyclic permutation under the matrix trace 
together with the former result (\ref{d+result}) yields
\begin{equation}
\label{spincont} 
\mtr \bigl( V_{T_{\ga_p^+}}\,d_{+,\ga_p^+}\,V_0^\dagger \bigr) = 
\mtr \bigl( V_0^\dagger\,V_0\,d_{+,\ga_p^+} \bigr) =
\mtr \big( d_{+,\ga_p^+} \bigr) =
2\,\cos(\tfrac{1}{2}\theta_{\ga_p^+})\,\cos\eta_{\ga_p^+} \ .
\end{equation}
Finishing the calculation as in the well known case of the Schr\"odinger 
equation \cite{Gut71,Gut90} finally yields a contribution of 
\begin{equation}
\label{pocont}
\chi(E)\,\frac{\tilde\varrho(T_{\ga_p^\pm})}{2\pi}\,A_{\ga_p^{\pm}}\,
\ue^{\frac{\ui}{\hbar}S_{\ga_p^\pm}(E)}\,\{1+O(\hbar)\}
\end{equation}
for every isolated non-degenerate periodic orbit $\ga_p^\pm$. Here the 
amplitude $A_{\ga_p^{\pm}}$ contains only classical information about 
the periodic orbit, including the contribution of a classical spin 
precessing along the orbit. Explicitly, the amplitude reads
\begin{equation}
A_{\ga_p^\pm} = \frac{2\,T^{\#}_{\ga_p^\pm}\,\cos(\tfrac{1}{2}
\theta_{\ga_p^\pm})\,\cos\eta_{\ga_p^\pm}}{\left|\det(M_{\ga_p^\pm} -
\eins_{4})\right|^{1/2}}\,\ue^{-\ui\frac{\pi}{2}\mu_{\ga_p^\pm}}\ .
\end{equation}
In this expression $T^{\#}_{\ga_p^\pm}$ denotes the primitive period of 
$\ga_p^\pm$, i.~e., the period of the associated primitive periodic orbit. 
Moreover, $M_{\ga_p^\pm}$ is the linearised Poincar\'e map (monodromy matrix) 
along the orbit and $\mu_{\ga_p^\pm}$ is its Maslov index. 

From the Gutzwiller trace formula it is well known that if $\ga_p^\pm$
is not primitive, all quantities that enter (\ref{pocont}), apart from the 
spin contribution (\ref{spincont}), can be readily expressed in terms of 
the respective quantities of the associated primitive periodic orbit. 
In order to extend this to (\ref{spincont}) we recall that if $\ga_p^\pm$ 
is a $k$-fold repetition of a primitive orbit, where $k\in\gz\setminus\{0\}$, 
the fact that $d_{+,\ga_p^+}$ is a holonomy implies that $d_{+,\ga_p^+}=
(d_{+,\ga_p^+}^{\#})^k$, where $d_{+,\ga_p^+}^{\#}$ denotes the 
corresponding primitive holonomy. According to \cite{Hor72} the trace 
(\ref{spincont}) associated with $\ga_p^\pm$ can hence be expressed as
\begin{equation}
\mtr \bigl( d_{+,\ga_p^+} \bigr) = T_k \left( \mtr \bigl( 
d_{+,\ga_p^+}^{\#} \bigr) \right) \ ,
\end{equation}
where $T_k (x)$ is the $k$-th Chebyshev polynomial of the second kind
in $x$. Thus, if desired, the following trace formulae (\ref{SemTF}) and 
(\ref{densTF}) can also be expressed in terms of sums over primitive 
periodic orbits and their repetitions.

In case the classical dynamics generated by the Hamiltonians $H^\pm$ have
only isolated non-degenerate periodic orbits, the relations (\ref{Weyl})
and (\ref{pocont}) now enable us to state the following semiclassical 
trace formula explicitly,
\begin{equation}
\label{SemTF}
\begin{split}
\sum_n \chi (E_n)\, \varrho \left( \frac{E_n -E}{\hbar} \right)
 & = \chi(E)\,\frac{\tilde\varrho (0)}{2\pi}\,\frac{2\,{\rm vol}\,(\Om_E^+) + 
   2\,{\rm vol}\,(\Om_E^-)}{(2\pi\hbar)^2}\,\left\{ 1+O(\hbar) \right\} \\
 & \quad +\sum_{\ga_p^{\pm}} \chi(E)\,\frac{\tilde\varrho(T_{\ga_p^\pm})}
   {2\pi}\,A_{\ga_p^{\pm}}\,\ue^{\frac{\ui}{\hbar}S_{\ga_p^\pm}(E)}\,
   \{1+O(\hbar)\} \ .
\end{split}
\end{equation}
The conditions imposed on the test function $\varrho$ and its Fourier
transform ensure that all expressions entering this trace formula are
finite. In particular, due to the compact support of $\tilde\varrho$ the 
sum over periodic orbits only includes orbits up to a finite period and 
since the Hamiltonian flow generated by $H^{\pm}(\vecp,\vecx)$ was supposed 
to be clean the sum is of finite length.

Often a semiclassical trace formula is presented for the spectral density
of the quantum Hamiltonian, see e.~g.\ \cite{Gut71,Gut90}. In the present
case one can readily obtain such a trace formula for the truncated
spectral density 
\begin{equation}
d_{\chi}(E) := \sum_n \chi(E_n)\, \delta(E-E_n) 
\end{equation}
from (\ref{SemTF}). This reads
\begin{equation}
\label{densTF}
\begin{split}
d_{\chi}(E) 
 & = \chi(E)\,\frac{2\,{\rm vol}(\Om_E^+) + 2\,{\rm vol}(\Om_E^-)}
     {(2\pi\hbar)^3} \, \left\{ 1 +O(\hbar) \right\} \\ 
 & \quad +\chi(E)\,\frac{1}{\pi\hbar}\,\sum_{\ga_p^{\pm}}
   \frac{T^{\#}_{\ga_p^{\pm}} \cos(\tfrac{1}{2}\theta_{\ga_p^{\pm}}) 
   \cos\eta_{\ga_p^{\pm}}}{\left| \det(M_{\ga_p^{\pm}} - \eins_4) 
   \right|^{1/2}} \, \ue^{\frac{\ui}{\hbar} S_{\ga_p^{\pm}}(E) 
   - \ui \frac{\pi}{2} \mu_{\gamma_p^{\pm}}} \left\{ 1+O(\hbar) \right\} \ .
\end{split}
\end{equation}       
Obviously, the sum over classical periodic orbits in (\ref{densTF})
does not converge. This trace formula rather has to be viewed as a 
distributional identity whose actual meaning is provided by (\ref{SemTF}).

%%%%%%%%%%%%%%%%% section 6 %%%%%%%%%%%%%%

\section{Nonrelativistic limit}

\label{sec6}

As compared to the Schr\"odinger equation, the Dirac equation takes
two generalisations into account. It first takes care of the spin degrees
of freedom and, secondly, describes relativistic dynamics. In many
situations of physical interest it, however, suffices only to include
spin and to leave the description of the translational motion on a  
nonrelativistic level. As it is well known, this can be achieved by 
considering the Pauli equation, possibly with several additional terms
such as one describing spin-orbit coupling. In this section we therefore 
want to study particles of charge $e$ and mass $m$ with spin 1/2 in a 
nonrelativistic context. To this end we compare the semiclassical
asymptotics for the Pauli equation with the leading order as 
$c\rto\infty$ of the time evolution kernel (\ref{Ksemicl}) for the 
Dirac equation. Both approaches, which will turn out to produce identical 
results, then allow to set up a semiclassical trace formula.

We recall that when one divides a Dirac four-spinor $\Psi$ into 
two-spinors $\psi_{1/2}$ according to 
\begin{equation}
\label{D+Pspinor}
  \Psi(\vecx,t) =: \left( \begin{matrix} \psi_1(\vecx,t) \\ 
                                         \psi_2(\vecx,t)
                   \end{matrix} \right) \ ,
\end{equation}
one obtains from (\ref{Diraceqn}) two coupled matrix differential 
equations. It is well known \cite{BjoDre64} that to leading order as 
$c \to \infty$ these equations decouple, and that for $\psi:=\psi_1$ the 
Pauli equation 
\begin{equation}
\label{Paulieqn}
  \ui \hbar \frac{\partial \psi}{\partial t} (\vecx,t) 
  = \left[ \frac{1}{2m} \left( \frac{\hbar}{\ui} \vecnab_{\vecx} - 
                               \frac{e}{c} \vecA(\vecx) \right)^2
           + e \, \varphi(\vecx) - \frac{e\hbar}{2mc} \, \vecsig \vecB(\vecx)
    \right] \psi(\vecx,t)  
  =: \hat{H}_P \psi(\vecx,t) 
\end{equation}
emerges. See also \cite{Tha92} for a careful treatment of the limit 
$c\rto\infty$. In Weyl quantisation the Pauli Hamiltonian $\hat{H}_P$ can 
be realised as 
\begin{equation}
  \hat{H}_P = H_P \left( \frac{\hbar}{\ui} \vecnab_{\vecx}, \vecx \right) \ ,
\end{equation}
with the $2\times 2$ symbol matrix
\begin{equation}
  H_P(\vecp,\vecx) = H_0(\vecp,\vecx) + \hbar\, H_1(\vecp,\vecx) \ .
\end{equation}
As opposed to the Dirac equation, compare (\ref{symbmat}), this Weyl
symbol is composed of a principal symbol 
\begin{equation}
\label{prinsymb}
  H_0(\vecp,\vecx) := 
  \left[ \frac{1}{2m} \left( \vecp - \frac{e}{c} \vecA(\vecx) \right)^2
  + e \, \varphi(\vecx) \right] \eins_2 \ ,
\end{equation}
which is a multiple of the identity matrix, and an additional subprincipal 
symbol
\begin{equation}
\label{subprinsymb}
  H_1(\vecp,\vecx) = - \frac{e}{2mc} \, \vecsig \vecB(\vecx) 
\end{equation}
that reflects a coupling of the spin degrees of freedom to the external
magnetic field.  

We are now going to study the semiclassical limit $\hbar\rto 0$ along
the lines developed above for the Dirac equation. Since therefore
many details of the following calculations are similar to the ones shown
in the previous sections, at several places the presentation will be kept 
rather brief. The matrix-valued Schwartz kernel $K_P(\vecx,\vecy,t)$ of 
the time evolution operator $\hat{U}_P(t) := \ue^{-\frac{\ui}{\hbar}
\hat{H}_Pt}$ obeys
\begin{equation}
  \psi(\vecx,t) = \int_{\rz^3} K_P(\vecx,\vecy,t) \, \psi_0(\vecy) 
                  \, \ud^3y \ , \quad
  \psi(\vecx,0) = \psi_0(\vecx) \ ,
\end{equation}
so that $K_P(\vecx,\vecy,t)$ has to solve the Pauli equation (\ref{Paulieqn}) 
with initial condition 
\begin{equation}
  \lim_{t \to 0+} K_P(\vecx,\vecy,t) = \eins_2 \, \delta(\vecx-\vecy) \ .
\end{equation}
Since the principal symbol $H_0$ is a scalar multiple of the identity
matrix and thus has one eigenvalue, only one Hamilton-Jacobi equation 
will be relevant to the semiclassical time evolution. We therefore choose 
the semiclassical ansatz 
\begin{equation}
\label{Pauliscan}
  K_P(\vecx,\vecy,t) = \frac{1}{(2\pi\hbar)^3} \int_{\rz^3}
  \left[ \sum_{k=0}^{\infty} \left(\frac{\hbar}{\ui}\right)^k 
  a_k(\vecx,\vecy,t;\vecxi) \,  
  \ue^{\frac{\ui}{\hbar}(S(\vecx,\vecxi,t)-\vecy\vecxi)}\right] \, 
  \ud^3\xi + O(\hbar^{\infty}) \ ,
\end{equation}
compare (\ref{oscillint}) and (\ref{genfct}), which we introduce into 
(\ref{Paulieqn}). Comparing like orders in $\hbar$ yields to lowest 
order a Hamilton-Jacobi equation for $S$, whereas to orders $\hbar^k$, 
$k=1,2,3,\dots$, transport equations for the $2\times 2$ matrices $a_{k-1}$
follow. The initial conditions for $S-\vecy\vecxi$ and $a_k$ are analogous 
to (\ref{phiini}) and (\ref{ampini}). By obvious reasons, the subprincipal 
symbol (\ref{subprinsymb}) cannot appear to zeroth order in $\hbar$ so 
that the Hamilton-Jacobi equation emerging in leading semiclassical order 
is only determined by the principal symbol $H_0(\vecp,\vecx)$. As a 
consequence, the classical dynamics of the translational degrees of freedom 
are those of a nonrelativistic point particle that does not experience a 
force coming from a coupling of spin to the external magnetic field. The 
latter is, however, contained in the transport equation for $a_0$ that 
appears in next-to-leading order. This reads 
\begin{equation}
\label{paulitransp}
  \left[ \vecnab_{\vecp} H_0(\vecnab_{\vecx}S,\vecx) \vecnab_{\vecx} 
         + \frac{\partial}{\partial t} + \frac{1}{2} \vecnab_{\vecx} 
         \left[ \vecnab_{\vecp} H_0(\vecnab_{\vecx}S,\vecx) \right] 
         + \ui M_P(\vecx) \right] a_0 = 0 \ ,
\end{equation}
so that the spin degrees of freedom enter through the traceless hermitian 
$2\times2$ matrix 
\begin{equation}
  M_P(\vecx) := - \frac{e}{2mc} \vecsig \vecB(\vecx) 
             \equiv H_1(\vecnab_{\vecx}S,\vecx) \ .
\end{equation}
In analogy to the considerations following (\ref{a0Sch}), the ansatz 
\begin{equation}
  a_0(\vecX(t),\vecy,t;\vecxi_j) = 
  \sqrt{ \det \left( 
  \frac{\partial^2 S}{\partial x_k \partial \xi_l}(\vecx,\vecxi_j,t) \right)}
  \, d_P(\vecP(t),\vecX(t)) \ ,
\end{equation}
for the lowest order amplitude, evaluated along a classical trajectory
$\ga_{xy}$ associated with the stationary point $\vecxi_j$ of the phase,
proves useful. It leads to the nonrelativistic spin transport equation 
\begin{equation}
\label{dPtransport}
  \left[ \frac{\ud}{\ud t} + \ui M_P(\vecX(t)) \right] d_P(\vecP(t),\vecX(t)) 
  = 0
  \ , \quad d_P(\vecP(0),\vecX(0)) = \eins_2 \ ,
\end{equation}
along $\ga_{xy}$ that determines the SU(2) matrix $d_{P,\ga_{xy}}$
describing the leading contribution of spin to the semiclassical time
evolution kernel. The transport equations (\ref{paulitransp}) and 
(\ref{dPtransport}) have already been obtained by Choquard \cite{Cho55},
who used a semiclassical ansatz similar to (\ref{Pauliscan}), however,
without the integration over $\vecxi$.

We remark that the decomposition of the SU(2) matrix $d_+$ into a classical
spin $\vecs$ and an additional phase $\eta$ presented in section \ref{sec4} 
can be repeated in the present, nonrelativistic, context. This procedure 
leads to the dynamical equation 
\begin{equation}
\label{nonrelBMT}
\frac{\ud}{\ud t}\vecs = \vecs \times \frac{e}{mc}\vecB
\end{equation}
for $\vecs$ transported along $\ga_{xy}$ in the external magnetic field. 
The r.h.s.\ of (\ref{nonrelBMT}) can readily be identified as the leading 
order of the r.h.s.\ of (\ref{BMT}) as $c\rto\infty$. An additional phase, 
\begin{equation}
\label{etaoftnr}
 \eta_P (t) = \frac{e}{2mc} \int_0^t \vecB(t') \vecs(t') \ \ud t'
              -\frac{1}{2} \int_0^t \big( 1 - \cos\theta(t') \big) 
              \,\frac{\ud\phi}{\ud t'}(t') \ \ud t' \ ,
\end{equation}
representing the third, nonclassical, degree of freedom of $d_P$, appears 
in the same manner as for the Dirac equation. This phase is again composed
of a dynamical and a geometric part.    

A calculation similar to that presented in section \ref{sec3} finally 
leads to the following semiclassical time evolution kernel for the Pauli 
equation (for $t\neq 0$),
\begin{equation}
\label{KscPauli}
  K_P(\vecx,\vecy,t) = \frac{1}{(2\pi\ui\hbar)^{3/2}} \sum_{\ga_{xy}}
    d_{P,\ga_{xy}} \, D_{\ga_{xy}} \, 
    \ue^{ \frac{\ui}{\hbar} R_{\ga_{xy}} 
          - \ui \frac{\pi}{2} \nu_{\ga_{xy}}} \{ 1+O(\hbar) \} \ .
\end{equation}
On the r.h.s.\ the sum extends over the solutions of Hamilton's equations 
of motion with the classical Hamiltonian $H_0(\vecp,\vecx)$, which connect 
$\vecy$ and $\vecx$ in time $t$. The corresponding Morse indices and 
Hamilton's principal functions are denoted by $\nu_{\ga_{xy}}$ and 
$R_{\ga_{xy}}(\vecx,\vecy,t)$, respectively; compare (\ref{Hamprifun}). 
The factor $D_{\ga_{xy}}$ is defined by 
\begin{equation}
  D_{\ga_{xy}} 
  := \left| \det \left( - \frac{\partial^2 R_{\ga_{xy}}}
     {\partial x_k \partial y_l} (\vecx,\vecy,t) \right) \right|^{1/2} \ .
\end{equation}
Since these quantities are already determined by the nonrelativistic 
classical dynamics generated by $H_0$, the only difference between  
(\ref{KscPauli}) and the respective semiclassical kernel for the 
Schr\"odinger equation is an appearence of the factors $d_{P,\ga_{xy}}\in
{\rm SU(2)}$ that represent the leading influence of spin. 

So far we have examined the propagator that emerges from first taking 
the nonrelativistic limit of the Dirac equation and then constructing a 
semiclassical time evolution kernel. We will now compare this to the 
result that one obtains by first constructing the relativistic semiclassical 
kernel (\ref{Ksemicl}) and then taking the nonrelativistic limit. We are 
thus now interested in the leading order behaviour of (\ref{Ksemicl}) as 
$c\to\infty$. To this end we first remark that 
\begin{equation}
  V(\vecp,\vecx) = \left( \begin{matrix} \eins_2 \\ 0 \end{matrix} \right) 
                   + O \left( \tfrac{1}{c} \right) \ , \quad
  W(\vecp,\vecx) = \left( \begin{matrix} 0 \\ \eins_2 \end{matrix} \right) 
                   + O \left( \tfrac{1}{c} \right) \ , \quad
  c \to \infty \ ,
\end{equation}
i.~e., in leading order the positive and negative kinetic energy parts in 
(\ref{Ksemicl}) decouple completely. We therefore have to compare 
(\ref{KscPauli}) with the upper left $2\times 2$ block of (\ref{Ksemicl}). 
This is consistent with (\ref{D+Pspinor}) because this block describes the 
time evolution of the two-spinor $\psi=\psi_1$. We furthermore recall the 
well known fact that the classical relativistic dynamics described by 
$H^+(\vecp,\vecx)$ turns into the nonrelativistic dynamics with the 
classical Hamiltonian $H_0(\vecp,\vecx)$ as $c \to \infty$. This in
particular implies that $\gamma_{xy}^+$, $R^+_{\gamma_{xy}^+}$, 
$\nu^+_{\gamma_{xy}^+}$ and $D^+_{\gamma_{xy}^+}$ may in leading order be 
approximated by $\gamma_{xy}$, $R_{\gamma_{xy}}$, $\nu_{\gamma_{xy}}$ and 
$D_{\gamma_{xy}}$. We are therefore only left with comparing the factors $d_+$ 
and $d_P$ which describe the influence of spin in a relativistic and in a 
nonrelativistic context, respectively. As already mentioned below equation
(\ref{nonrelBMT}) 
\begin{equation}
  M_+ = M_P + O(\tfrac{1}{c}) \ , \quad c\to\infty \ ,
\end{equation}
leads to a nonrelativistic approximation for both the classical spin
dynamics and the nonclassical phase. Thus $d_P$ provides the leading order
asymptotical term for $d_+$ as $c\rto\infty$. Collecting everything
one observes that in the nonrelativistic limit the upper left block of 
the semiclassical time evolution kernel (\ref{Ksemicl}) for the Dirac 
equation turns into the respective result (\ref{KscPauli}) for the Pauli 
equation. In this sense the limits $\hbar\rto 0$ and $c\rto\infty$ 
commute, at least concerning leading orders.

In order to set up a semiclassical trace formula for the Pauli Hamiltonian
we now assume that $\hat H_P$ has a pure point spectrum. Otherwise we
would have to employ an energy localisation to a gap in the essential
spectrum as described in section \ref{sec2} for the Dirac Hamiltonian. 
We then consider a test function $\varrho\in\cS(\rz)$ with compactly 
supported Fourier transform $\tilde\varrho$. Applying the procedure of 
section \ref{sec5} to $\hat H_P$ finally yields
\begin{equation}
\label{PauliTF}
\begin{split}
\sum_n \varrho \left( \frac{E_n -E}{\hbar} \right) 
 &=\frac{\tilde\varrho(0)}{2\pi}\,\frac{2\,{\rm vol}(\Om_E)}
  {(2\pi\hbar)^2}\,\{ 1+O(\hbar) \} \\ 
 &\quad +\sum_{\ga_p}\frac{\tilde\varrho(T_{\ga_p})}{2\pi}\,A_{\ga_p}\,
  \ue^{\frac{\ui}{\hbar}S_{\ga_p}(E)}\, \{ 1+O(\hbar) \} \ ,
\end{split}
\end{equation}
where the sum extends over the periodic orbits of the classical dynamics
generated by the principal symbol $H_0$. In fact, all classical quantities 
entering (\ref{PauliTF}) refer to this Hamiltonian. As explained in the 
relativistic case, for such a trace formula to be valid the flow generated 
by $H_0$ must be clean. The amplitude associated with each isolated,
non-degenerate periodic orbit then reads
\begin{equation}
A_{\ga_p} = \frac{2\,T^{\#}_{\ga_p}\,\cos(\tfrac{1}{2}\theta_{\ga_p})\,
\cos\eta_{P,\ga_p}}{\left|\det(M_{\ga_p} - \eins_{4})\right|^{1/2}}\,
\ue^{-\ui\frac{\pi}{2}\mu_{\ga_p}}\ .
\end{equation}

A different semiclassical approach to the Pauli equation has previously
been used to investigate spin-orbit coupling \cite{LitFly92,FriGuh93}.
The authors of these papers principally base their method on the technique 
developed 
by Littlejohn and Flynn \cite{LitFly91a,LitFly91b} to treat matrix-valued 
wave operators with principal symbols that have no (globally) degenerate 
eigenvalues. Below we will derive the time evolution according to the 
prescription of the semiclassical limit employed in \cite{LitFly92,FriGuh93}, 
however, using the techniques outlined in the previous sections. For 
simplicity, and for ease of comparison with our previous semiclassical 
study of the Pauli equation, we will not consider spin-orbit coupling but 
only a coupling of spin to the external magnetic field. Therefore, the 
relevant Hamiltonian is the one defined in (\ref{Paulieqn}). However, a 
generalisation of the following discussion to arbitrary couplings of the 
form $\vecsig\,\vecC(\vecp,\vecx)$ is straight forward. Following now 
the philosophy of \cite{LitFly92,FriGuh93}, we introduce Bohr's magneton 
$\mu:=\frac{e\hbar}{2mc}$ and consider it as constant in the semiclassical 
limit. Thus the Hamiltonian
\begin{equation}
\hat {H'}_P := \frac{1}{2m} \left( \frac{\hbar}{\ui} \vecnab - \frac{e}{c} 
               \vecA(\vecx) \right)^2 + e \, \varphi(\vecx) - 
               \mu\, \vecsig \vecB(\vecx)
\end{equation}
arises as a Weyl operator associated with the symbol
\begin{equation}
\label{modsymb}
{H'}_P (\vecp,\vecx) := \frac{1}{2m} \left( \vecp - \frac{e}{c} \vecA(\vecx) 
                        \right)^2+ e \, \varphi(\vecx) - \mu\, 
                        \vecsig \vecB(\vecx) \ ,
\end{equation}
so that no subprincipal symbol occurs. As opposed to the situation 
analysed at the beginning of this section, one could view the present
procedure as taking the simultaneous limits $\hbar\rto 0$ and $|\vecB|
\rto\infty$ in such a way that $\hbar|\vecB|=const$. Another way to
look at this is to keep $|\vecB|$ fixed, but to perform the limit of
`large spin'. 

As long as $\vecB\neq 0$ the $2\times 2$ symbol matrix (\ref{modsymb})
has two non-degenerate eigenvalues
\begin{equation}
\label{modHam}
{H'}_P^\pm (\vecp,\vecx) = 
  \frac{1}{2m} \left( \vecp - \frac{e}{c} \vecA(\vecx) \right)^2
  + e \, \varphi(\vecx) \mp \mu\, |\vecB (\vecx)| \ ,
\end{equation}
which generate two classes of classical dynamics. Since this is similar
to the situation occurring for the Dirac equation, the semiclassical
ansatz for the time evolution kernel $K'_P(\vecx,\vecy,t)$ therefore
should be chosen as in (\ref{oscillint}), but where now the amplitudes
are $2\times 2$ matrices. Applying the procedure described in section
\ref{sec3} then leads to two Hamilton-Jacobi equations, with the two
Hamiltonians (\ref{modHam}).

The non-scalar contribution to the symbol matrix (\ref{modsymb}) is
given by the Hamiltonian $-\mu\,\vecsig\vecB(\vecx)$ describing a quantum 
mechanical precessing spin. The orthonormal eigenvectors $v_\pm (\vecx)
\in\kz^2$ of ${H'}_P (\vecp,\vecx)$ corresponding to the eigenvalues 
${H'}_P^\pm(\vecp,\vecx)$ are hence well known from the standard example 
of the Berry phase \cite{Ber84}. In analogy to (\ref{a0pm}) one can now 
introduce the ansatz
\begin{equation}
a_0^+ (\vecx,\vecy,t;\vecxi) = 
v_+ (\vecx)\,\tilde v^\dagger(\vecx,\vecxi,t)
\end{equation}
with some suitable vector $\tilde v\in\kz^2$. Upon multiplication of the 
transport equation for $a_0^+$ with $v_+^\dagger (\vecx)$ from the left 
one obtains the equation
\begin{equation}
\left[ \vecnab_{\vecp}{H'}_P^+ (\vecnab_{\vecx}S^+,\vecx) \vecnab_{\vecx} 
         + \frac{\partial}{\partial t} + \frac{1}{2} \vecnab_{\vecx} 
         \left[ \vecnab_{\vecp}{H'}_P^+ (\vecnab_{\vecx}S^+,\vecx) \right] 
         + \ui {M'}_P^+ (\vecnab_{\vecx}S^+,\vecx) \right] 
         \tilde v^\dagger = 0 \ ,
\end{equation}
where
\begin{equation}
{M'}_P^+ (\vecp,\vecx) := -\frac{\ui}{m} \left( \vecp - \frac{e}{c} 
                          \vecA(\vecx) \right) \, v_+^\dagger (\vecx)
                          \vecnab_{\vecx}v_+ (\vecx) \ .
\end{equation}
On the r.h.s., the factor $v_+^\dagger\vecnab v_+$ is the well known 
expression for the adiabatic connection leading to the U(1)-Berry phase 
of a precessing spin \cite{Ber84}. Again, the transport equations for 
$a_0^\pm$ will be solved along the classical trajectories ${\ga'}^\pm_{xy}$ 
following from the Hamiltonians (\ref{modHam}). As suggested by (\ref{D+def}) 
and (\ref{d+def}), one separates
\begin{equation}
\tilde v^\dagger (\vecX(t),\vecxi_j,t) = 
  \sqrt{\det\left(\frac{\partial^2 S^+}{\partial x_k\partial\xi_l}
  (\vecX(t),\vecxi_j,t) \right)}\ {d'}_+ (\vecP(t),\vecX(t))\ v_+^\dagger
  (\vecy)\ ,
\end{equation}
so that along a trajectory ${\ga'}^+_{xy}$ the phase 
${d'}_+(t)\in{\rm U(1)}$ has to solve the equation
\begin{equation}
\label{Berryph}
\left[ \frac{\ud}{\ud t} + v_+^\dagger (\vecX(t)) \vecnab_{\vecx} 
v_+ (\vecX(t))\,\dot{\vecX}(t) \right] {d'}_+(t) = 0\ , 
\qquad {d'}_+(0) = 1 \ .
\end{equation}
As a result, one obtains the well known U(1)-Berry phase of a quantum 
mechanical spin that has been transported adiabatically along the 
trajectory ${\ga'}^+_{xy}$. Finally, the semiclassical time evolution kernel
reads
\begin{equation}
\label{KPmod}
{K'}_P (\vecx,\vecy,t) 
  = \frac{1}{(2\pi\ui\hbar)^{3/2}} \sum_{{\ga'}^\pm_{xy}}v_\pm(\vecx)
    v_\pm^\dagger(\vecy)\,{d'}_\pm\,D^\pm_{{\ga'}^\pm_{xy}}\,
    \ue^{\frac{\ui}{\hbar}{R'}^\pm_{{\ga'}^\pm_{xy}}-\ui\frac{\pi}{2}
    \nu_{{\ga'}^\pm_{xy}}}\ \{ 1+O(\hbar) \} \ ,
\end{equation}
where all classical quantities refer to the dynamics generated by the
Hamiltonians (\ref{modHam}) and are defined in analogy to the previous 
cases. In particular, Hamilton's principal functions 
${R'}^{\pm}_{{\ga'}^{\pm}_{xy}}(\vecx,\vecy,t)$ are related to the analogous 
quantities $R_{{\ga'}^{\pm}_{xy}}(\vecx,\vecy,t)$ that are defined by the 
Hamiltonian (\ref{prinsymb}), but here evaluated along the trajectories
${\ga'}^{\pm}_{xy}$, through
\begin{equation}
\label{RandR'}
{R'}^\pm_{{\ga'}^\pm_{xy}}(\vecx,\vecy,t) = 
R_{{\ga'}^\pm_{xy}}(\vecx,\vecy,t) \pm \mu\int_0^t |\vecB(\vecX(t'))|
\ \ud t' \ .
\end{equation}
Due to the factors $v_\pm(\vecx)v_\pm^\dagger(\vecy)$ in (\ref{KPmod})
the `spin up' and `spin down' components of an initial spinor 
$\psi_0(\vecy)$ are propagated independently along the corresponding
trajectories following from the Hamiltonians ${H'}^\pm_P$. Here `spin up' 
and `spin down' are defined with respect to the instantaneous direction
of the magnetic field. This procedure breaks down at mode conversion
points, i.~e., at points where the magnetic field vanishes. There the
symbol matrix (\ref{modsymb}) has one twofold degenerate eigenvalue and
the level surfaces of the two Hamiltonians ${H'}^\pm_P$ cross. Moreover,
since $\vecnab_{\vecx}{H'}^\pm_P$ develops a singularity, the classical
trajectories are not smooth when crossing a mode conversion point. In
certain situations this defect can be cured by letting the trajectories
cross the two level surfaces, see \cite{FriGuh93}, but in general an 
application of the present semiclassical procedure requires a more 
refined treatment of mode conversion points, see \cite{LitWei93} for
a detailed discussion. 

At this place one could easily establish the corresponding semiclassical 
trace formula, if one followed the programme outlined in section \ref{sec2} 
once again. However, we refrain from doing this here and rather comment 
on the relation between the two semiclassical approaches to the Pauli 
equation discussed in this chapter, which lead to the two distinct 
expressions (\ref{KscPauli}) and (\ref{KPmod}) for the time evolution 
kernel. In the first scenario we systematically performed an expansion 
in $\hbar$ and determined the leading order terms for the time evolution
kernel and for the classical side of the trace formula. We observed that
to lowest order the translational degrees of freedom decouple from the 
spin degrees of freedom in that the translational motion experiences no
back reaction from the coupling of spin to the external magnetic field.
In a certain sense this decoupling can be seen as an adiabatic one where 
the translational motion is considered as slow, although this condition is not 
needed for the formulae to be valid. This topic was also discussed by Balian 
and Bloch \cite{BalBlo74} in the context of semiclassical approximations for
the Green function. Spin enters in 
next-to-leading order and, among other quantities, determines the 
amplitudes in the relevant semiclassical expressions. The leading order 
of the spin dynamics is given by that of a classical spin precessing along 
the particle trajectories. We have repeatedly emphasized that at this stage 
no adiabatic limit is considered. In addition, a geometric phase of the type 
discussed by Aharonov and Anandan \cite{AhaAna87} enters the amplitudes.
In the second scenario we considered the double limit $\hbar\rto 0$,
$|\vecB|\rto\infty$, $\hbar|\vecB|=const$. It turned out that in this
context the relevant classical translational motion follows from
two Hamiltonians, taking the effect of a coupling of a `spin up'
and a `spin down', respectively, to the external magnetic field into
account. Since in this context, via the expressions $v_\pm(\vecx)
v_\pm^\dagger(\vecy)$ in (\ref{KPmod}), the spin direction is defined with 
respect to the instantaneous direction of the magnetic field, the spin 
degrees of freedom are transported adiabatically along the particle 
trajectories. There is no further dynamical equation for a classical spin 
and, as a consequence of the limit $|\vecB|\rto\infty$, the Berry phase 
that enters can be viewed as emerging from an adiabatic approximation of the 
geometric term found in the first scenario. This finding is in agreement 
with the remark on the relation of these phases that can be found in 
\cite{AhaAna87}. In order to be more specific concerning the issue of 
adiabaticity, we introduce spherical coordinates $(\vt_B,\vp_B)$ for 
$\vecB/|\vecB|$ which allows to obtain an explicit expression for the 
Berry phase emerging from (\ref{Berryph}). Moreover, we reintroduce 
$\frac{e\hbar}{2mc}$ for the magneton $\mu$. Using (\ref{RandR'}) one 
thus observes that
\begin{equation}
{d'}_\pm\,\ue^{\frac{\ui}{\hbar}{R'}^\pm_{{\ga'}^\pm_{xy}}} = 
\exp\left\{\pm\ui\frac{e}{2mc}\int_0^t|\vecB|\,\ud t' \mp\frac{\ui}{2}
\int_0^t (1-\cos\vt_B)\,\dot{\vp}_B\,\ud t' \right\}\,\ue^{\frac{\ui}{\hbar}
R_{{\ga'}^\pm_{xy}}}\ ,
\end{equation}
which can be introduced in (\ref{KPmod}). The first factor on the r.h.s.\ 
is readily identified to arise as an adiabatic approximation to 
$\ue^{\pm\ui\eta_P}$, if the phase (\ref{etaoftnr}) is evaluated along the 
trajectory ${\ga'}^+_{xy}$ instead of $\ga_{xy}$. Thus the spin contributions
to (\ref{KPmod}) are adiabatic approximations to the respective contributions
to (\ref{KscPauli}). Conversely, one concludes that in the semiclassical 
time evolution according to (\ref{KscPauli}) the spin transport is performed 
non-adiabatically so that mode conversion poses no difficulty. In
conclusion, one can consider the two alternative ways of performing the
semiclassical limit as being considered with a `weak' and a `strong' 
coupling, respectively, of spin to the translational degrees of freedom. 
Comparing the two results (\ref{KscPauli}) and (\ref{KPmod}) for the time 
evolution kernel one notices that the two approaches yield different 
results when extended to intermediate couplings. In the first case
the translational motion is not influenced by the spin, whose dynamics in
turn is not treated adiabatically. In the second case, however, spin has
an effect on the translational motion, but the spin dynamics enters in
an adiabatic approximation. This observation illustrates the fact that 
the $\hbar$-expansions employed are not uniform in the field strength.  

\subsection*{Acknowledgement}
We would like to thank Roman Schubert for useful discussions.

%%%%%%%%%% references %%%%%%%%%%%%%

\end{document}